\documentclass[journal]{IEEEtran}

\usepackage{graphicx}
\usepackage{subfig}
\usepackage{amsmath,amsbsy,amsgen,amscd,amsthm,amsfonts,amssymb} 
\usepackage[mathscr]{euscript}
\usepackage{centernot}
\usepackage{tikz}
\usepackage{bbold}

\usetikzlibrary{bayesnet}
\usetikzlibrary{arrows}
\usetikzlibrary{backgrounds}

\newcommand{\assign}{:=}

\DeclareMathOperator*{\argmin}{arg\,min}

\begin{document}

\title{Explainable AI for a No-Teardown Vehicle Component Cost Estimation:\\ A Top-Down Approach}

\author{Ayman~Moawad,
        Ehsan~Islam,
        Namdoo~Kim,
        Ram~Vijayagopal,
        Aymeric~Rousseau,
        and~Wei~Biao~Wu% <-this % stops a space
\thanks{A. Moawad is with the Vehicle and Mobility Simulation group at Argonne National Laboratory, 9700 S Cass Ave, Lemont, IL 60439 USA and the department of Statistics, The University of Chicago, 5801 S Ellis Ave, Chicago, IL 60637 USA. E-mail: amoawad@anl.gov, aymoawad@uchicago.edu}% <-this % stops a space
\thanks{Ehsan Islam, Namdoo Kim, Ram Vijayagopal and Aymeric Rousseau, are with the Vehicle and Mobility Simulation group at Argonne National Laboratory.}% <-this % stops a space
\thanks{Wei Biao Wu is with the department of Statistics at The University of Chicago.}}

\maketitle

\begin{abstract}
The broader ambition of this article is to popularize an approach for the fair distribution of the quantity of a system's output to its subsystems, while allowing for underlying complex subsystem level interactions. Particularly, we present a data-driven approach to vehicle price modeling and its component price estimation by leveraging a combination of concepts from machine learning and game theory. We show an alternative to common teardown methodologies and surveying approaches for component and vehicle price estimation at the manufacturer's suggested retail price (MSRP) level that has the advantage of bypassing the uncertainties involved in 1) the gathering of teardown data, 2) the need to perform expensive and biased surveying, and 3) the need to perform retail price equivalent (RPE) or indirect cost multiplier (ICM) adjustments to mark up direct manufacturing costs to MSRP. This novel exercise not only provides accurate pricing of the technologies at the customer level, but also shows the, a priori known, large gaps in pricing strategies between manufacturers, vehicle sizes, classes, market segments, and other factors. There is also clear synergism or interaction between the price of certain technologies and other specifications present in the same vehicle. Those (unsurprising) results are indication that old methods of manufacturer-level component costing, aggregation, and the application of a flat and rigid RPE or ICM adjustment factor should be carefully examined. The findings are based on an extensive database, developed by Argonne National Laboratory, that includes more than 64,000 vehicles covering MY1990 to MY2020 over hundreds of vehicle specs.
\end{abstract}


% \fi
\IEEEpeerreviewmaketitle

\section{Introduction}

\IEEEPARstart{D}{eveloping} vehicle component models and  building physics-based vehicle models is a common approach among engineers in the automotive industry. Car manufacturers and researchers in the field have been able to develop a series of tools and simulation-based processes that evaluate the effects of advanced vehicle technologies on energy consumption. For more than two decades, Argonne National Laboratory has supported the U.S. Department of Energy (DOE) Vehicle Technologies Office (VTO) Analysis Program by estimating the impact of new technologies on the energy consumption and cost of several thousand vehicles \cite{moawad_assessment_2016}, \cite{islam_extensive_2018}. To estimate the overall impact, the VTO's Analysis group sponsors different vehicle market penetration tools that rely on Argonne’s vehicle energy efficiency and cost estimates. Although vehicle energy models have been continuously developed and validated with test data, the uncertainty surrounding vehicle cost estimation has been increasing, with the latest studies being several years old \cite{ricardo_autonomie_2010}.\par
Vehicle pricing depends directly on the vehicle's attributes, the powertrain-related components' power and size, and the materials used, as well as the manufacturing complexity, volume, manufacturer's reputation and marketing strategies. As the name suggests, the manufacturer's suggested retail price (MSRP) is the recommended selling price calculated by the manufacturer's financial experts in order to earn a competitive rate of return on its investments in technology. It covers direct costs, such as costs of materials and labor, but also indirect costs, such as costs associated with R\&D, pensions and other employee benefits, warranties, advertising, and manufacturer and dealer profits. Thus the MSRP is appropriate measure to study to understand vehicle price evolution over time as well as the distribution of vehicle price over technology. It also reflects the price paid by consumers in competitive market conditions, which is relevant for evaluating the costs and benefits of fuel economy and the resulting market penetration impacts, and can also be used to calculate the per-vehicle cost increase of Corporate Average Fuel Economy (CAFE) rules.\par
Previous methodologies tended to approach the problem of vehicle price estimation with a bottom-up approach \cite{ricardo_autonomie_2010}, \cite{hill_improving_2016}, \cite{lutsey_efficiency_2017}. Essentially, a vehicle teardown analysis is performed based on a limited number of high-volume/high-sales vehicle data points, from which a series of technology cost curves are developed. These cost equations are then used in aggregation to estimate the total vehicle manufacturing cost, and a fixed retail price equivalent (RPE) methodology is used to mark up direct manufacturing costs to MSRP. \par
When the model year (MY) 2012-2016 greenhouse gas and CAFE standards were developed for the 2011 Average Fuel Economy Standards Passenger Cars and Light Trucks Model Final Rule  \cite{nhtsa_average_2009}, \cite{rogozhin_automobile_2009}, DOE used an RPE of 1.5 in conjunction other indirect cost multipliers (ICM), which resulted in an average markup of 1.25 \cite{whinihan_retail_2012}. Ricardo, an environmental consulting services company with which Argonne worked when developing the standards \cite{ricardo_autonomie_2010}, suggested that indirect cost ``must be contained within an external markup factor, either an RPE factor, typically 1.5, or an ICM, which varies from 1.02 to 1.45 depending on the technology complexity and time frame of interest.'' The National Research Council, on the other hand, found that a RPE of 2.0 is more adequate. That being said, this flat approach has also been open to criticism as different levels of profitability across vehicle classes or product lines are recognized among cost analysts.\par
Ricardo's work in supporting the development of component cost models for Argonne was based on transactional component prices from an independent supplier to a vehicle manufacturer and includes costs associated with the manufacture and development of the component, system integration costs, vehicle assembly costs, vehicle manufacturer and dealer selling expense, and margins. This effort let to the development of 10 technology module cost models: low voltage system, engine system, engine accessories, transmission system, vehicle drivetrain system, energy storage system, e-drive system, fuel cell system, hydraulic propulsion system, and glider system. Costs were valued in 2010 dollars.\par
The authors are not aware of any recent publicly available study that attempts to update the acquired cost curves based on up-to-date vehicle data. In addition, there are concerns about the limited number of data points available to the entities developing the cost estimates. The collected vehicle data misses to reflect all the novel technologies implemented nowadays (e.g., cylinder deactivation or Skip-Fire engines, 10-speed transmissions, etc.). It is also important to consider the inherent interactions between vehicle components and their effect on vehicle price. For example, although individual components may have fixed manufacturing and labor costs, a combination of several advanced technologies can potentially be packaged by the manufacturer at a different price point not necessarily related to the cost of manufacturing. The studies described earlier fail to address this kind of interaction, and so does a applied fixed RPE estimate. Earlier studies also missed to address the issue of correlated features within a vehicle: The presence of a certain advanced technology in a sub-part of the vehicle increases the chances of including advanced technologies in other parts of the vehicle. For example, advanced turbo engines are likely to be found in vehicles with advanced transmissions with a high number of gears, or the higher the engine power of the vehicle the more unlikely it is to find basic/elementary technologies or attributes in other aspects of the vehicle. Cost curves developed in isolation run the risk of misrepresenting the resulting aggregated total vehicle price.

\section{Contribution}
As described in the introduction, in general the basic method of cost estimation is to tear down technologies within a carefully selected series of vehicles and construct a bottom-up estimate by costing out material, labor and manufacturing processes. An alternative method is to acquire estimates of selling prices of manufactured components. Both methods are rather tedious, expensive, and rely on a certain level of expertise on the part of the estimator (who carry some level of bias). In addition, many original equipment manufacturers (OEMs) manufacture their own components and maintain a strict level of confidentiality. Information about the prices of component parts acquired by third-party suppliers is also not easily accessible, although there have been several useful reports \cite{nrc_effectiveness_2002}, \cite{nescaf_reducing_2004}. Information can also be obtained from the confidential data manufacturers submit and share with governmental institutions or from discussions with OEMs or suppliers. Other attempts have been made in the past to estimate component price by comparing the prices of vehicles with and without the technology or component of interest \cite{duleep_analysis_2008}.\par

In this article, we propose to take a top-down approach ― from vehicle retail price (MSRP) estimation using machine learning techniques down to component price attributional effects on MSRP ― by leveraging game theory concepts of coalition and cooperation \cite{shapley_value_1953}. At the very least, the authors are seeking to popularize the use of a novel and delightful alternative methodology within the community and encourage all to improve upon it.\par

In the following sections we will detail the different efforts undertaken during this study. Significant efforts have been made to:
\begin{itemize}
  \item Collect a large and reliable amount of vehicle data, with a detailed level of specification and technology breakdown. As a result, Argonne has exclusively developed an internal Vehicle Attribute Database (ArVAD) that includes more than 64,000 vehicles from MY1990 to MY2020. ArVAD contains hundreds of vehicle features: vehicle MSRP, color, front and rear seat details, vehicle measurement details, drivetrain information, fuel-related information, engine specs and technologies, power feature details (such as power or heated mirrors, remote keyless power, etc.), vehicle comfort details, instrumentation information, vehicle entertainment packages, tire and wheel specifications, suspension technologies, etc.
  \item Cluster the vehicles: to automate the categorization of vehicles into baseline, performance, luxury, and prestige\footnote{Unique and specially manufactured vehicles.} regardless of the manufacturer's name or reputation, the model or the trim level. This clustering step has been found vital to vehicle price modeling, as it reduces model variance and increases model accuracy. In fact, in a model based on powertrain attributes only, for example, price variability for vehicles with similar powertrain specifications can be large, depending on the  manufacturer's car line category (standard vs. luxury) as well as non-powertrain-related specifications such as the presence of other advanced options. A classic example of this kind is the price discrepancy between some Honda and Acura vehicles when a very limited number of differences (if any) can be observed.
  \item Analyze outcome interpretability, and fine-tune several machine learning models, settling on a state-of-the-art gradient boosting on a decision trees algorithm called Catboost \cite{prokhorenkova_catboost:_2019}, \cite{dorogush_catboost_2018}.
  \item Understand, interpret, and explain model outcomes and predictions and how they relate to the vehicle  features/technologies input. This article will discuss and detail the use of a feature attribution method based on the computation of Shapley values, a method from coalitional game theory \cite{shapley_value_1953}. In particular, this article attempts to popularize a framework for optimal credit allocation and the explanation of individual predictions \cite{lundberg_unified_2017}, \cite{lundberg_explainable_2019}.
  \item Describe and analyze the marginal effects that vehicle components have on the total vehicle price. As a result of this analysis, the authors develop and suggest a ``non-equation'' based method for vehicle price estimation and component price attribution. We call the methodology proposed Shapley-based credit/penalty component pricing (SCP). We will show that this penalty approach can be used to assess the costs and benefits of fuel economy, including such activities performed for U.S. regulatory analysis. In particular, this novel methodology can help regulatory entities evaluate the incremental cost of increasing vehicle efficiency.
\end{itemize}

\section{Purpose and Potential Beneficiaries}
The primary, direct purpose of this research activity was to support the U.S. DOE VTO Analysis Program and explore a potential novel approach to update current vehicle and component price estimation methods involved in the various benefits analysis studies conducted. VTO relies on the Argonne-developed software environment Autonomie\footnote{Autonomie is a MATLAB-based software environment and framework for automotive control system design, simulation, and analysis. The tool is designed for rapid and easy integration of models with varying levels of detail and abstraction as well as processes. Developed by Argonne National Laboratory in collaboration with General Motors, Autonomie was designed to serve as a single tool that can be used to meet the requirements of automotive engineering throughout the development process, from modeling to control. Autonomie is used to evaluate the energy consumption and cost of advanced powertrain technologies. It has been validated for several powertrain configurations and vehicle classes using Argonne's Advanced Powertrain Research Facility vehicle test data. Autonomie is the primary vehicle simulation tool selected by U.S. DOE to support its U.S. Drive Program and Vehicle Technologies Office. It has been used in numerous studies to provide the U.S. government with guidance for future research. More than 175 companies and research entities, including major automotive companies and suppliers, use Autonomie to support their advanced vehicle development programs.} to handle vehicle energy and price estimation efforts that feed into subsequent market penetration tools. Other entities can benefit from the outcome of this work, particularly other governmental and regulatory entities that evaluate the incremental cost of increasing vehicle efficiency, or manufacturers that perhaps would like to advance their understanding of product and component pricing among competitors.

\section{Collection Process and Data}
This project takes a data-driven approach, and therefore its success depends on the richness and quality of the data in hand. For that  reason, Argonne has expended significant work to develop an internal vehicle attribute database by leveraging web-scraping techniques to collect publicly available data.
The research team focused especially on developing a general, automated data collection and web-scraping process to collect vehicle data. The process allows researchers to efficiently crawl the web by deploying a web spider that targets car and OEM websites. The web-scraping framework contains four modules that control the  process:

\begin{enumerate}
  \item \textbf{The spider module.} Defines what we want to extract from the web page of interest.
  \item \textbf{The request/response module.} Handles the request sent to the website and the content of that request through the injection of custom headers and assignment of proxies, then manages the download of the data received from the website response.
  \item \textbf{The processing module.} Takes care of cleaning the data, removing duplication, and storing it in the appropriate form and data structure.
  \item \textbf{The manager module.} Responsible for preserving operation orders and the priorities of scheduled requests. It coordinates among all the pieces for consistent operation while accounting for website response delays, lagging, and multiple simultaneous requests.
\end{enumerate}

Images, vehicle specifications and other publicly available information (including vehicle MSRP) is fetched and stored in a non-relational database (MongoDB), resulting in an exhaustive dataset that can be used to build a precise vehicle MSRP estimation model.\par

Argonne completed several data processing steps in building the database:
\begin{itemize}
  \item \textbf{Cleaning.} Data have been checked for missing values and inconsistencies.
  \item \textbf{Integration.} Data from various sources have been successfully integrated into a large dataset.
  \item	\textbf{Modification.} Outliers have been identified and fixed using cross-references of the different sources and imputation methods available.
  \item \textbf{Transformation and feature engineering.} Several additional calculated fields were created.
  \item \textbf{Analysis and interpretation.} Several rounds of data analysis were performed.
\end{itemize}

The database contains an extensive list of vehicle features: power and energy specifications, drivetrain information, measurements, instrumentation, interior and exterior options, entertainment components (such as sound systems/speakers, screens, and other things that can affect vehicle pricing), and detailed information about tires and wheel specifications (type, width, aspect ratio, diameter, load index, speed rating, etc.).
VTO's objective is to construct a model in which the MSRP estimation is driven primarily by powertrain components rather than luxury features. However, to reduce model variance and uncertainty, some non-powertrain features are included in the modeling, and basic/standard attributes will be used as input for predictions to reflect Autonomie's standard/average vehicle segments.\par
\begin{figure}[!t]
\centering
\includegraphics[width=2.5in]{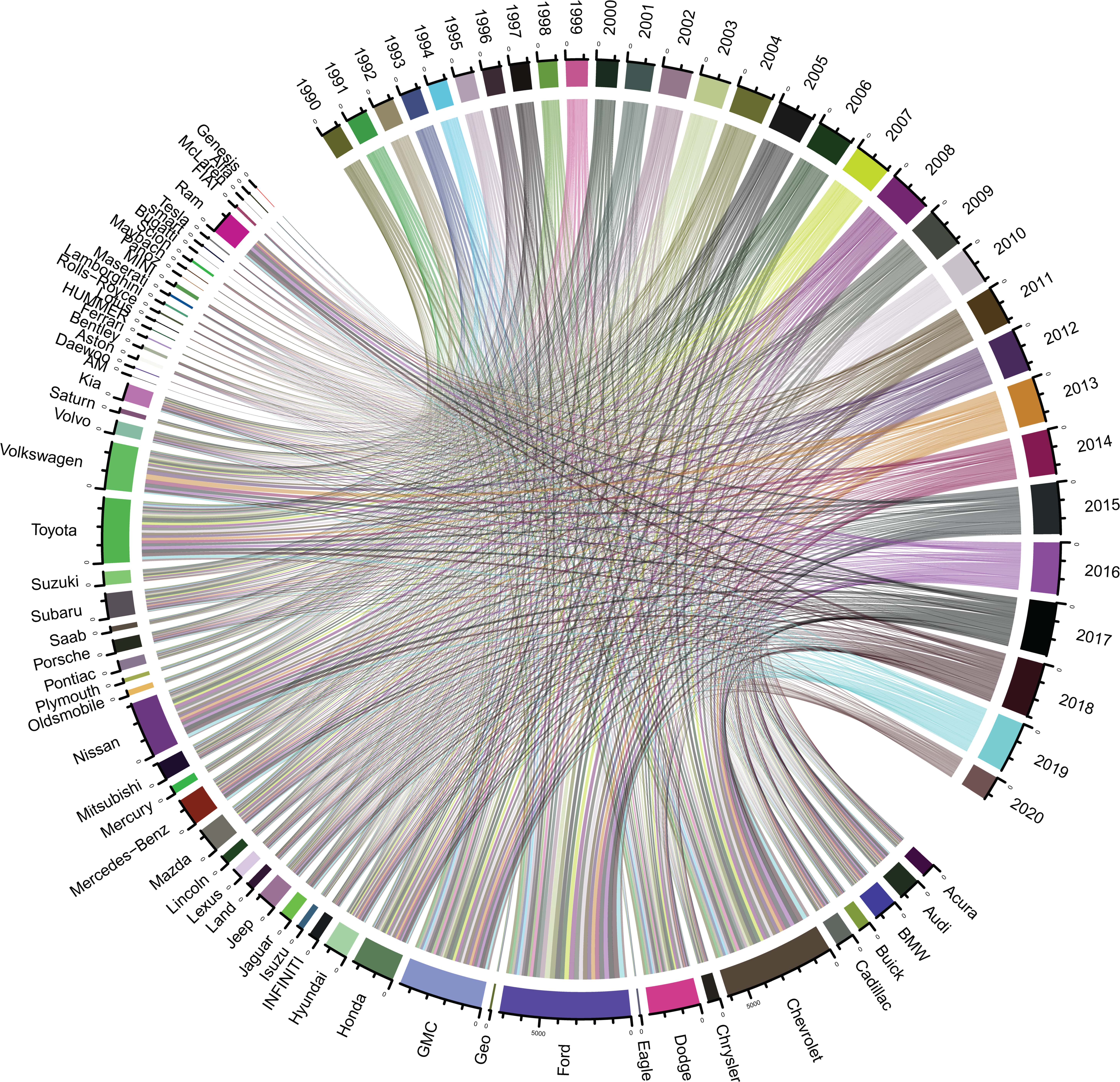}
\caption{Chord plot showing count relationship of vehicle make and vehicle year.}
\label{fig:chord}
\end{figure}

The dataset currently includes some 64,000 vehicles, from 1990 to 2020,\footnote{As of 12/2019. Web crawling is performed on a monthly basis to update the database with newly appearing models.} of various makes, models, and trim levels with hundreds of variables/specs. The data exhaustively cover many vehicle make manufacturers from 1990 to the present (figure \ref{fig:chord}), and we note a general uniform distribution of makes over the years for big and established OEMs. Some newer companies, such as Tesla, will need special treatment during modeling due to the unique technologies they exhibit in terms of powertrain type, specs and others (e.g., electric powertrain, navigation systems, etc.).
At this writing,  the data collected on MY2020 vehicles were still limited, and many models were not yet released.

About 152 variables have been selected for analysis. A variable selection study has been performed to carefully select features, assess their importance, understand their  degrees of correlation with MSRP, and assess the explanatory power of each variable. Note that the dataset displays a mix of variable types: Some variables are numeric (e.g., engine power), other are categorical (e.g., transmission type), or more specifically Boolean (e.g., engine has turbocharging technology T/F).\par
\begin{figure}[!t]
\centering
\includegraphics[width=2.5in]{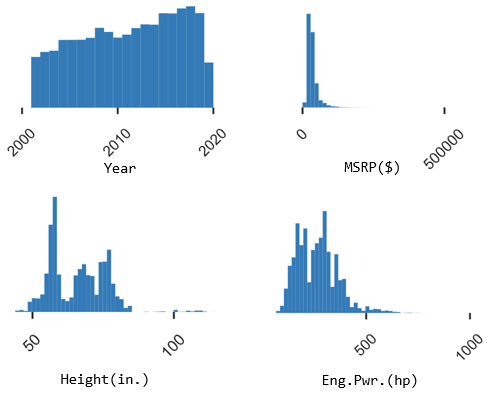}
\caption{Snapshot of distribution exhibited by some of the variables in the data.}
\label{fig:snapshot}
\end{figure}
\begin{figure*}[!t]
\centering
\subfloat[Correlation heatmap of numerical features.]{\includegraphics[width=2.5in]{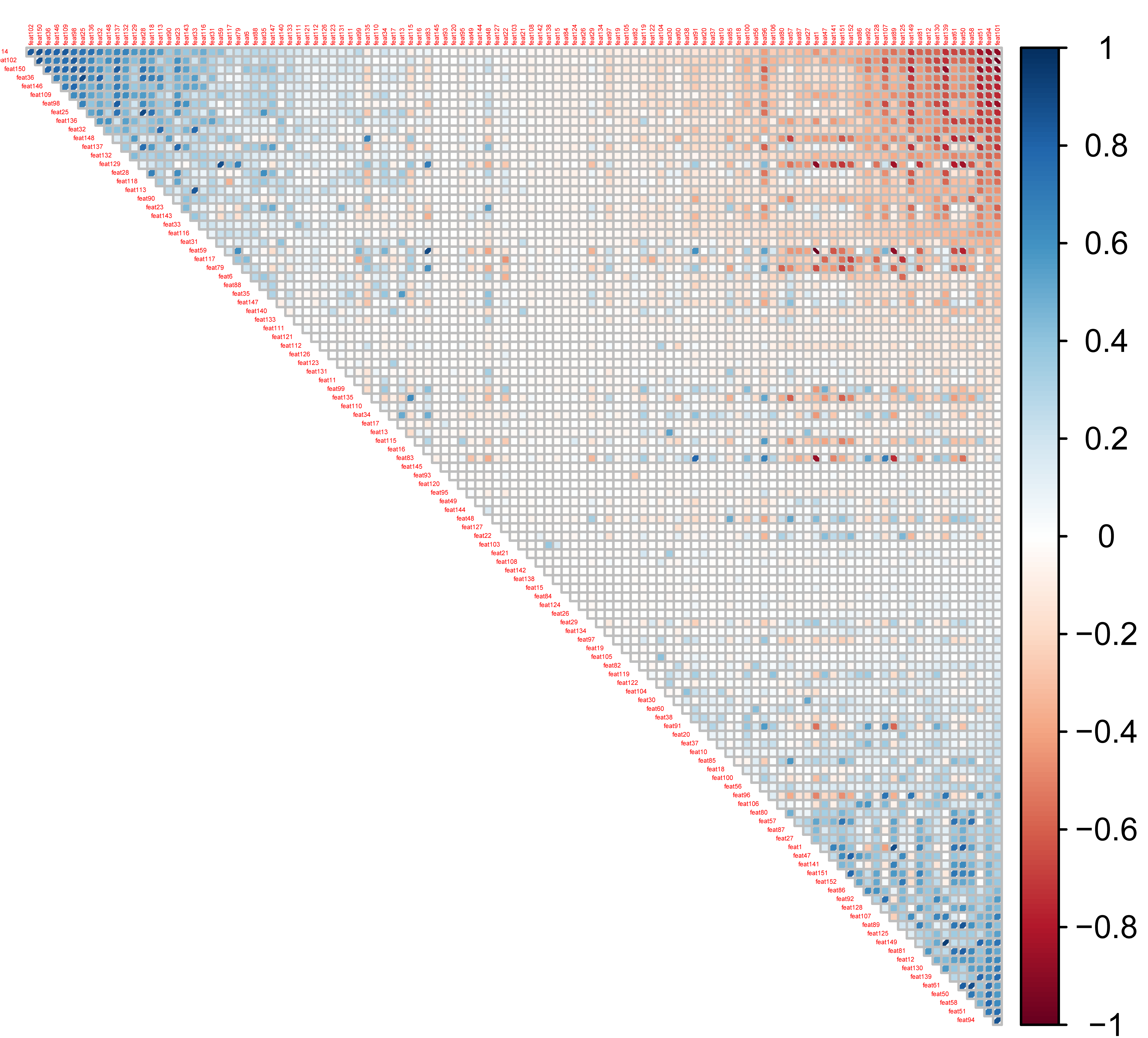}%
\label{fig:corra}}
\hfil
\subfloat[Subset (zoom)]{\includegraphics[width=2.5in]{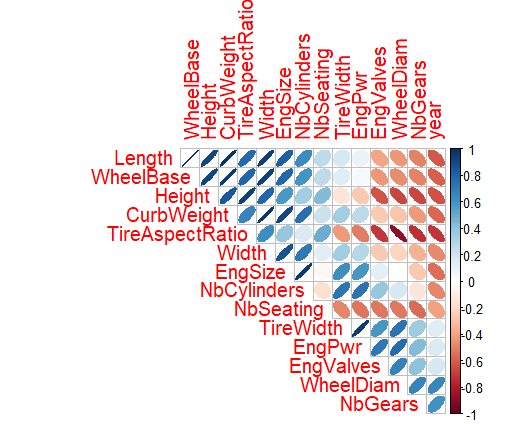}%
\label{fig:corrb}}
\caption{Correlogram.}
\label{fig:corr}
\end{figure*}

Figures \ref{fig:snapshot} and \ref{fig:corr} show a glimpse of the underlying distributions and existing correlations for some of the variables. We note that there is a increase in the number of  models appearing every year. Several other interesting and not unexpected facts arise: Vehicle MSRP distribution has a clear heavy right tail, with most vehicles prices being in the \$0-100,000 range. The mean and median for this distribution are respectively $\sim$\$34,000 and \$29,000, and the data exposes quite a large vehicle price variance as well. There is an apparent "multi" mode of engine power in the distribution; this information coupled with vehicle curb weight can give us an idea of the different clusters of power density values existing in the data, and this, along with vehicle dimensions,  can be used as proxy for vehicle classification. The next section will discuss the creation of vehicle clusters to reduce model variance during the modeling phase. Finally, we note from the correlogram certain groups of variables with strong positive or negative correlations. For example, engine power and acceleration are strongly positively correlated.

Keeping in mind that the purpose of the current modeling is to ``model'' vehicle prices and extract component price values as well, we considered, as part of the variable selection process, the removal of systemic non-causal variables. As in this example, the investigation of the causal impact of certain variables has been carefully distinguished. Here, acceleration or vehicle performance is a causal descendant of other system related variables (power, weight, etc.), and consequently the correlation with MSRP can be largely explained by those parent variables.

\section{Make and Model Agnostic Clustering}
As noted above, DOE is interested in estimating vehicle segments related to the baseline segment. This is in line with Autonomie practice and its vehicle models, which represent the average market vehicle for each powertrain. To segregate base, luxury, performance, and prestige vehicles for proper modeling without knowledge of the make, model or trim level, a clustering approach is needed.\par

Several clustering algorithms were considered, but the \textit{interpretable} hierarchical clustering method gave good results. The hierarchical clustering method groups data points using a bottom-up approach (agglomerative) based on selected features as a measure of similarity. The agglomerative approach in hierarchical clustering is an important and well-established technique in unsupervised machine learning, where the clustering algorithm starts from singleton nodes (vehicles) and iteratively merges pairs based on mutual closeness. The process is repeated until all vehicles have been aggregated into one mega-cluster. Throughout the process, the merging blueprint is recorded and later revealed in the form of a dendrogram from which we have the flexibility to select an adequate number of clusters, segregating the vehicles according to our needs. This clustering approach requires careful selection of distance metrics as well as a measure of inter-cluster dissimilarity. For more detail, there is extensive literature on the subject \cite{murtagh_survey_1983}, \cite{reddy_survey_2018}.\par

In our setting, the main assumption driving our clustering is that vehicles of comparable size, performance, and other carefully selected specifications (e.g., vehicle weight, wheel radius) should be comparable in price, and therefore should be clustered together. The effect of this assumption is that inter-cluster vehicles with significant price differences represent different car lines (e.g., luxury).\par

\begin{figure}[!t]
    \centering
    \includegraphics[width=2.5in]{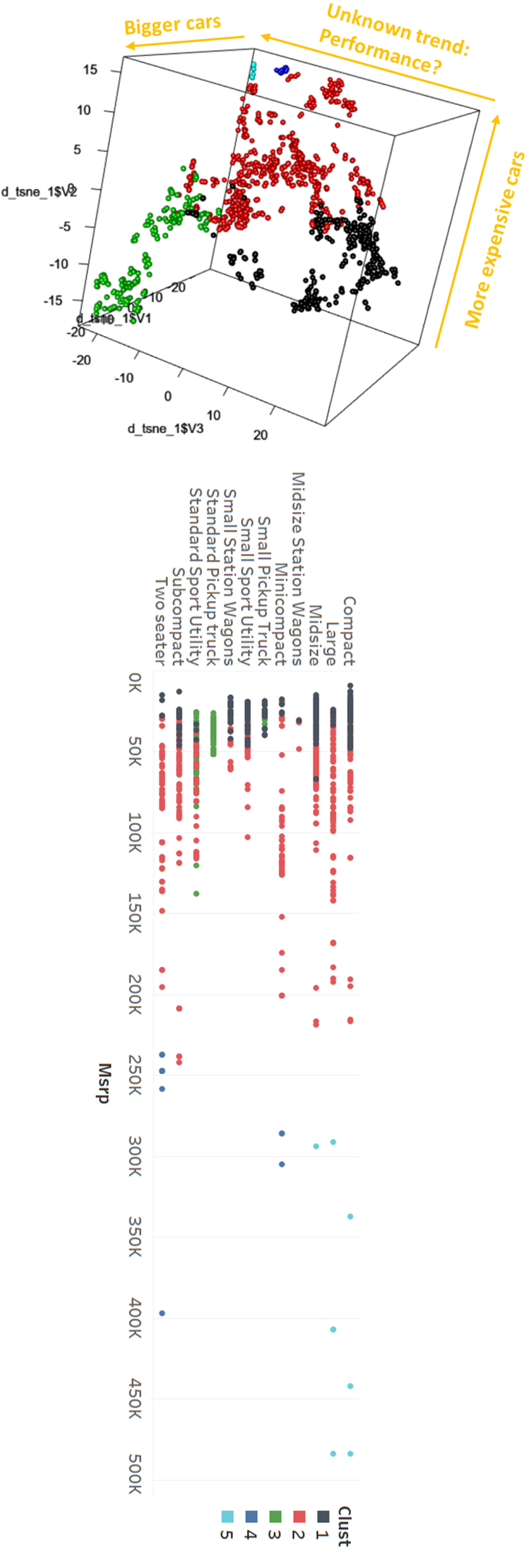}
    \caption{Vehicles clusters (Clust) t-SNE 3D projection (top) and a per class interpretation against the associated MSRP (bottom)}
    \label{fig:clustering}
\end{figure}

The advantage of the hierarchical clustering method is the ability to visualize the resulting tree-based division using a dendrogram to facilitate interpretation. In addition, there is some theoretical support for an optimal number of cluster choices, a task that is always difficult to achieve in unsupervised clustering algorithms. Figure \ref{fig:clustering} shows the resulting clustering projected onto a three-dimensional space. This is achieved by using the t-distributed stochastic neighbor embedding (t-SNE) dimensionality reduction technique (right). t-SNE is considered the current state-of-the-art dimension reduction technique that can produce a low dimensional representation of high dimensional data while preserving local distance structures \cite{maaten_visualizing_2008}, \cite{van_der_maaten_accelerating_2014}. This visualization allows us to cross-check the behavior of the resulting clustering. In fact, the projection shown provides clues about the interpretation of the results. The yellow axes describe the authors' best guess of the clustering interpretation after a quite extensive analysis. The green cluster is separated due to a clear differentiation in vehicle dimension specifications. The red cluster seems to represent luxury car lines, while the black cluster suggest baseline vehicles. A few vehicles, in light and dark blue, distinctly belong to more prestigious categories.\par

The bottom plot in figure \ref{fig:clustering} identifies clusters by vehicle class, which reveals additional details. Within each class, there is a clear separation between base and luxury vehicles (cluster 1 versus cluster 2). Interestingly, a third cluster, for larger vehicles (pickups), emerges. Where no within-class discrimination is apparent, this seems to support the fact that all pickups usually belong to one car line (a clear small variance in MSRP for pickups is also seen). Clusters 4 and 5 represent the most expensive vehicles, which are eliminated from the analysis and dataset, they can be considered to be exceptional outlier vehicles skewing the data (roughly above \$250k). Vehicles from clusters 1 and 3 are combined into one to represent base vehicles ($\sim$\$0–\$80,000) of all class types. Cluster 2 represents the luxury car line ($\sim$\$30,000–\$240,000). This method allows a soft price margin for vehicle segment segregation, so there is an overlap.\par
As noted earlier, the clustering preparation provides additional information to the modeling phase to reduce variance and increase explainability.

\section{Vehicle Price Model}
The vehicle price modeling approach taken in this work falls into a typical discriminative supervised learning setting. Given a dataset $\mathscr{D}=\{(\boldsymbol{X_i},y_i)\}_{i=1,\ldots n}$ of $n$ pair of examples consisting of a vector of $\boldsymbol{X_i} \in \mathbb{R}^m$ explanatory variables and $y$, a response or output variable, we want to learn a function $f:\boldsymbol{X} \mapsto y$ that can predict $y^*$ for new or unobserved or future inputs $\boldsymbol{X^*}$. In the following, $\boldsymbol{X}$ will refer to a carefully selected set of vehicle attributes, chosen according to the rules of explanatory power, as described in the previous sections, but also conforming to the engineering sense and the domain knowledge of the authors. The variable $y$ will refer to the vehicle price output (MSRP).\par
While it is assumed that the data $\mathscr{D}$ is sampled from some unknown distribution $p(\boldsymbol{X},y)$, we are not concerned with learning the distribution from available data. In the following we will detail how $f \in \mathscr{F}$ is chosen from the function space $\mathscr{F}$ of decision trees ― the hypothesis space ― to minimize $\mathbb{E}_{(X,y) \sim p}L(f(X),y)$ for the typical squared loss $L$.

\subsection{Catboost Model}
Catboost model is a state-of-the art machine learning model, based on gradient boosting, with a novel successful handling of categorical features. Gradient boosting on decision tree algorithms is very popular for problems with heterogeneous features in tabular form. Those algorithms are designed to achieve competitive results in the presence of complex, noisy and highly feature-dependent data \cite{chen_xgboost_2016}, \cite{ke_lightgbm_2017}.\par
The Catboost algorithm has the advantage of overcoming categorical data pre-processing, which typically involves some form of naive transformation of the data. One hot encoding, i.e., adding a binary feature as indicator for the category, is one approach \cite{miccibarreca_preprocessing_2001}, but high cardinality leads to infeasible processing and training. Other approaches have been considered to limit the number of features generated, such as grouping categories by target statistics (TS)  \cite{miccibarreca_preprocessing_2001}, which estimates the target expected value in each category. That is, if we are given a dataset $\mathscr{D}=\{(\boldsymbol{X_i},y_i)\}_{i=1,\ldots n}$, where $\boldsymbol{X_i}=(x_{i,1},\ldots,x_{i,m}) \in \mathbb{R}^m$ is a vector of $m$ features, possibly categorical, with $y_i \in \mathbb{R}$, then $x_{i,k}$ is substituted by: $$x_{i,k}=\frac{\sum_{j=1}^n \mathbb{1}_{\{x_{j,k}=x_{i,k}\}}.y_j}{\sum_{j=1}^n \mathbb{1}_{\{x_{j,k}=x_{i,k}\}}}$$
Other approaches convert categorical variables into gradient numerical statistics \cite{noauthor_features_nodate}.\par
The estimation just described can be noisy, especially for low frequency categories. Catboost uses an approach based on the performance of a random permutation of the dataset, after which the average target value is calculated for the example based on the same categorical values placed before the permutation. That is, if $\sigma = (\sigma_1, \ldots, \sigma_n)$ is a permutation, then $x_{\sigma_p,k}$ is replaced with: $$x_{\sigma_p,k}=\frac{\sum_{j=1}^{p-1} \mathbb{1}_{\{x_{\sigma_j,k}=x_{\sigma_p,k}\}}.y_{\sigma_j}+a.P}{\sum_{j=1}^{p-1} \mathbb{1}_{\{x_{\sigma_j,k}=x_{\sigma_p,k}\}}+a}$$ where $P$ is a prior value and $a$ a weight parameter imposed on the prior value \cite{cestnik_estimating_1990}. \par
The $P$ value can simply be set as the average response value of the dataset. This smoothing manipulation allows Catboost to overcome overfitting problems, but it also allows the use of the whole dataset for training in an efficient online manner. The introduction of random permutation for the purpose of calculating target statistics  is a strategy against data leakage (target leakage), in which a new feature $x_{i,k}$ is computed from $y_k$ but overcomes conditional shift \cite{zhang_domain_2013} when the distribution of $\boldsymbol{X_i}|y$ in the training set differs from the test set. This is a typical problem of learner generalization error  that Catboost is addressing innovatively.\par
Gradient boosting models assume that the data $\mathscr{D}$ is sampled from some unknown distribution $p(\boldsymbol{X},y)$, then given a loss function $L:\mathbb{R}^2 \rightarrow \mathbb{R}_+$. The goal is to find a function $F: X \rightarrow \mathbb{R}$ that minimizes the empirical risk: $$\mathscr{L(F)}=\sum_{i=1}^n L(F(\boldsymbol{X_i}),y_i)$$ such that: $$F(\boldsymbol{X})=\sum_{k=1}^t \alpha f_k(\boldsymbol{X})$$ where $t$ is the number of iterations. Typically, for best results $f_k \in \mathscr{F}$ is chosen from the space $\mathscr{F}$ of decision trees functions \cite{breiman_classification_1984}, \cite{friedman_additive_2000}. In other words, each of the $t$ functions $f_k$ is an independent tree structure separating the feature space $\mathbb{R}^m$ into several disjoint regions based on the value of a splitting feature.\footnote{Catboost makes use of oblivious trees \cite{langley_oblivious_1994}, \cite{kohavi_oblivious_1995}, \cite{ferov_enhancing_2016}, \cite{lou_bdt_2017}} Those $f_k$ functions are called base or weak learners and are learned sequentially by constructing the sequence $f_1,\ldots,f_t$ such that $$f_t = \argmin_{f \in \mathscr{F}} \mathscr{L}(F_{t-1}+f)=\argmin_{f \in \mathscr{F}} \sum_{i=1}^n L(F_{t-1}(\boldsymbol{X_i})+f(\boldsymbol{X_i}),y_i)$$
There are several ways to perform this optimization problem. Some are based on first order derivative calculations of $\mathscr{L(F)}$ at point $F_{t-1}$ and use the gradient as the step of minimization in a gradient descent type of optimization setting (i.e., using least squares):

\begin{equation}\label{eq:gb}
  f_t = \argmin_{f \in \mathscr{F}} \sum_{i=1}^n \Big(f(\boldsymbol{X_i})+\frac{\partial L(\hat y_i,y_i)}{\partial y_i}\Bigr|_{\substack{\hat y_i=F_{t-1}(\boldsymbol{X_i})}}\Big)^2
\end{equation}

We have presented only a quick description of the most basic gradient-based models. Additional details can be found in  other references, e.g., \cite{friedman_greedy_2001}, in which the learning objective can be regularized to treat overfitting problems, and stochastic gradient boosting can be applied \cite{friedman_stochastic_2002}  to improve the quality of the learner.\par
Catboost leverages gradient boosting to approximate gradient implementation; however, Catboost integrates several tricks to address an apparent weakness in gradient estimation \cite{breiman_outbag_1996}, \cite{breiman_using_2001}: The quantity in (\ref{eq:gb}) is biased due to the bias of the point-wise gradient estimates, because the conditional distribution of the gradient $\frac{\partial L(\hat y_i,y_i)}{\partial y_i}\Bigr|_{\substack{\hat y_i=F_{t-1}(\boldsymbol{X_i})}}$ for a given $\boldsymbol{X_i}$ is shifted from that of a test set. To overcome this problem, Catboost proposes an ordered boosting algorithm detailed in \cite{prokhorenkova_catboost:_2019} and \cite{dorogush_catboost_2018} that does not suffer from prediction shift.\par

\subsection{Model Performance}
It is worth mentioning that before settling on the Catboost algorithm, we tested several other ensemble learning models: AdaBoost, XgBoost, LightGBM, and Random Forest. We also used and analyzed standard (multilayer perceptron) neural network, support vector machine regression models, and Bayesian networks \cite{fernandez_extension_2008} and assessed them for out-of-sample performance. Simpler interpretable linear regression based models (fine-tuned with added complexity, interactions and regularization) have also been developed for optimal prediction performance while attempting to preserve explainability. Careful hyperparameter tuning and configuration selection (when applicable) with distributed grid search has been carried out throughout each exercise. In the end, Catboost outperformed all the models considered.\par
Catboost training was performed on NVIDIA TITAN Xp and NVIDIA Quadro P2000 graphics processing units (GPUs). The training time was on the order of several hours ($\sim 10^3$ minutes). The final ensemble model consists of $\sim 4000$ trees of depth=4, i.e., allowing for order-4 levels of interactions. All model parameters and hyperparameters were tuned in line with nested cross-validation methods for training and testing, using a validation set of vehicles separate from the test set for final performance assessment. The data was randomly split into five 80/20 folds for training and testing. Out-of-sample prediction performance was evaluated on all outer test sets. An inner loop then addresses, for each training set, a 80/20 split for hyperparameter tuning and calibration. For result stability, we bootstrapped over 100 iterations.\par
\begin{figure}[!t]
\centering
\subfloat[Root mean squared error, mean absolute percentage error, explained variance and residuals of averaged five-fold cross-validation.]{\includegraphics[width=2in]{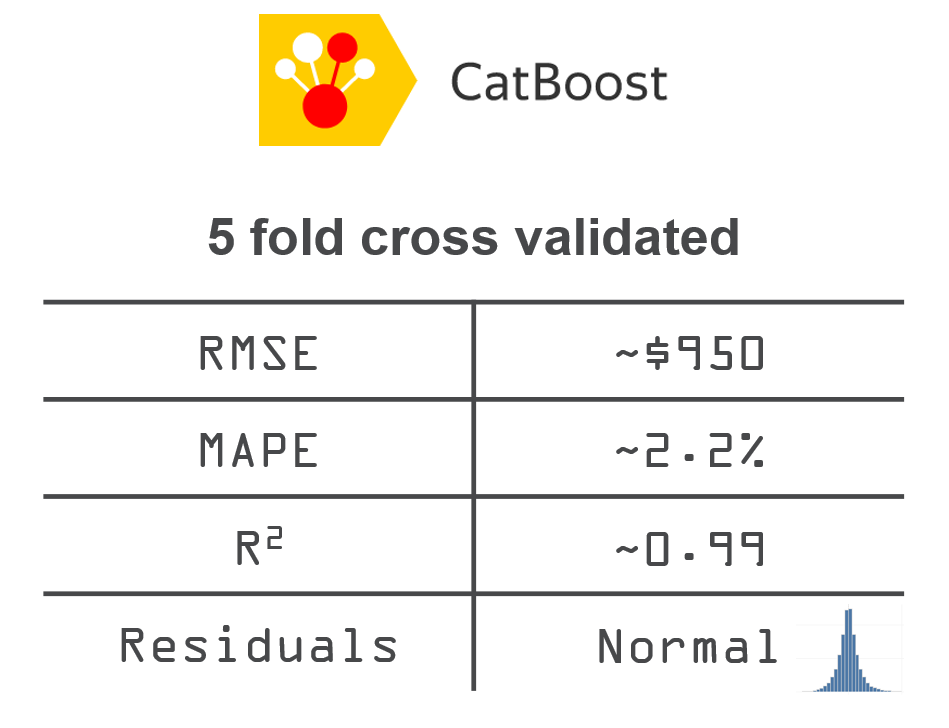}%
\label{fig:perf}}
\\
\subfloat[Prediction vs. true vehicle MSRP for \$0-100,000 vehicles.]{\includegraphics[width=2.5in]{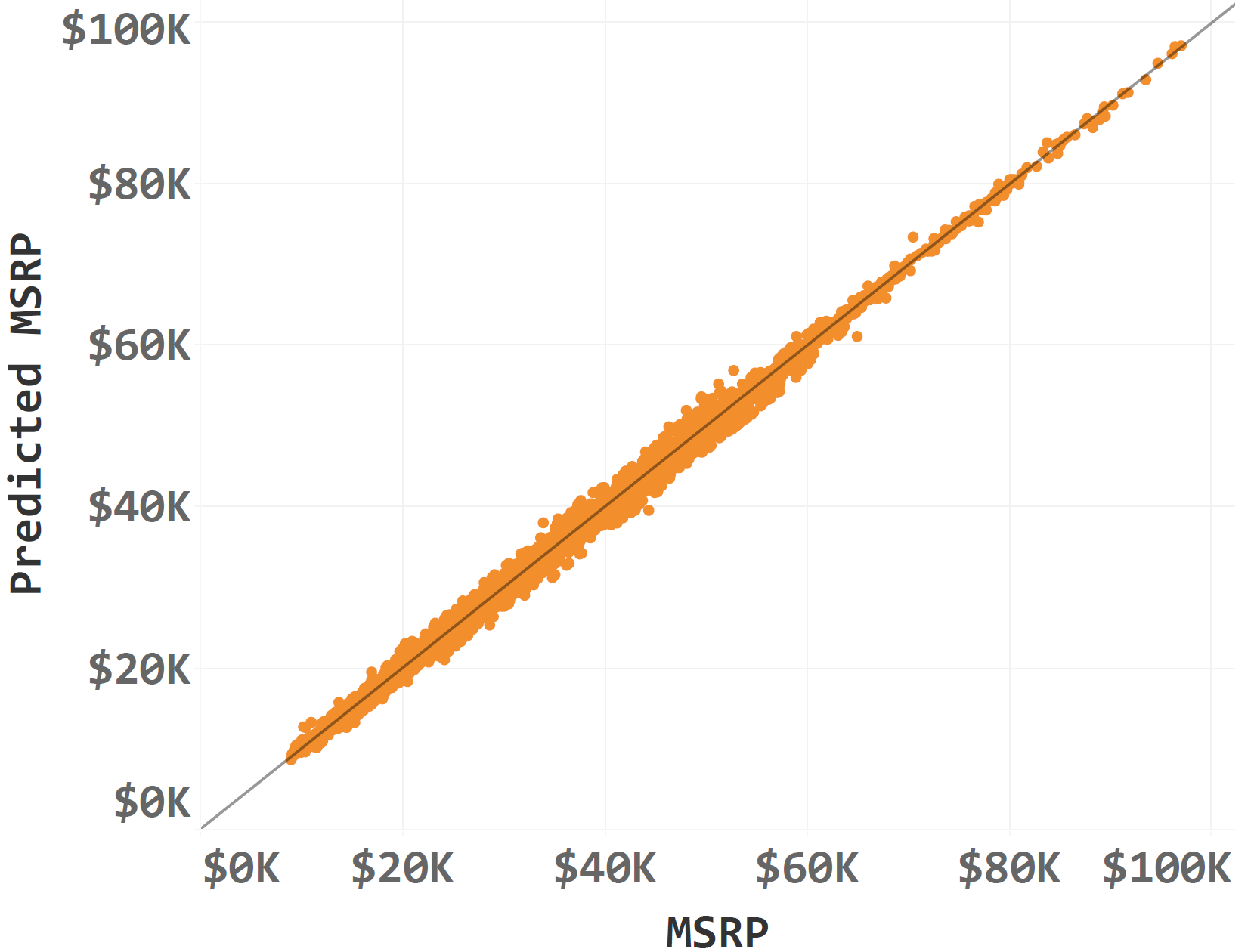}%
\label{fig:pred}}
\caption{Model Performance.}
\end{figure}

The model performance summary in figure \ref{fig:perf} shows a comprehensive recap of some of the five-fold cross-validated performance metrics.

The root mean squared error (RMSE) $\assign \sqrt{\frac{1}{n}\sum_i (y_i-\hat y_i)^2}$ shows the average vehicle MSRP prediction error a little less than \$1000, corresponding to a 2.2\% average error computed by the mean absolute percentage error (MAPE) $\assign \frac{1}{n}\sum_i |\frac{y_i-\hat y_i}{y_i}|$ on the test sets. It can be said that vehicle MSRPs are predicted with very reasonable precision. The normal shape of the residuals centered at zero suggests that most vehicle price predictions are within a few hundreds dollars of the MSRP. Figure \ref{fig:pred} shows the predicted vs. actual vehicle price for \$0-100,000 vehicles where there is a notably thin cloud of prediction points around the line $y=x$, hence the high $R^2$ value.

\subsection{Residual Analysis}
From an engineering perspective, uniform prediction accuracy levels is required. For generalization purposes, the modeling ignores manufacturer specificity. As a result, it is necessary to ensure that the resulting model accuracy is not biased towards a certain group, type or class of vehicle.\par
We performed a series of residual plots that analyze the behavior of the prediction error for a selected attribute of interest. We put special attention to analyze the distributions of residuals over manufacturers, vehicle type and classification. Also, because of the temporal dependency present in the data we tested for correlated errors through a combination of simple successive pairs of residuals plotting and Durbin-Watson statistic tests. We also considered more complex dependencies, although not likely in the vehicle pricing setting: no short or long times series runs of residuals was identified. Normality was checked, and residual behavior against specifically selected predictor variables was inspected carefully. For example, we confirmed homoscedasticity of the residual plot over the vehicle engine power and weight to guarantee that the model performs well across a wide range of vehicles in size and performance. Overall, we insured that residual properties that analysts would want to have in such diagnostics are showing satisfactory behavior, suggesting a good overall fit.

\section{Vehicle Component Price Estimation}
A we have shown, the  Catboost model developed has excellent prediction accuracy, but unfortunately is very opaque due to the complexity of the underlying gradient boosting based structure. Catboost, like many other complex machine learning models, is very flexible, accounting for many layers of nonlinear and interactional structures. This makes the statistical inference, a crucial requirement for the current application, challenging. For the purpose of extracting component price estimates from the predicted total vehicle price values, a certain level of model interpretability is required. Given the model output $f(x_1,\ldots,x_m)$, one would want to quantify to what extent each $x_j$ is responsible for the output value. In other words, with such a complex model, the challenge is to find a way to account for how the input features relate to the predictions without loss of model accuracy. A lot of good work has been done to perform the inverse exercise, which attempts to build simpler (although sometimes complex and sophisticated) but carefully designed models that enable us to explain as well as estimate the effect that each of the input components has on the response. Typically, for example, in linear regression models, model coefficients describe how a unit change in an explanatory variable affects a model response change, while holding other variables constant ― a sometimes impossible task.\par
Other methods employ post-hoc model-agnostic interpretation methods, such as the partial dependence plots (PDP) proposed in \cite{friedman_greedy_2001} or individual conditional expectation (ICE) \cite{goldstein_peeking_2013}, to explain complex models, but they can produce misleading results. PDP and ICE can give biased results when high degrees of feature codependence exist, a very common situation, and interactional behaviors are not well captured or quantified.\par
We turned to a game theory method to quantify to what extent each component contributes to vehicle price prediction and to retrieve individual component pricing.

\subsection{Shapley Method}
\label{sec:shapmeth}
A promising recent contribution to interpretable machine learning has emerged for proper feature attribution in non-linear complex settings \cite{lundberg_unified_2017}, \cite{lundberg_explainable_2019} \cite{lundberg_consistent_2019}. The work presented here is based on coalitional game theory methods using the computation of Shapley values \cite{shapley_value_1953}. The basic idea was originally developed by the economist Lloyd Shapley while he was working on the problem of fairly allocating credits among players in a game of cooperating players. The method has been adapted from the original purpose ― fairly allocating credit for the outcome of a game to collaborating players who may have contributed unequally ― to the purpose of fairly allocating credit to features of a model for the output of that model. The Shapley approach has the advantage of having strong theoretical support to ensure a fair feature attribution and consequently a fair distribution of the total prediction value among the features and their individual contributions.\par

The explanation of complex models via Shapley values starts by defining a class of additive feature attribution methods that will be used as a surrogate explanation model for the original one. If $f$ is the original prediction model and $g$ is an explanation model, then an additive feature attribution model is a linear function of binary variables in the form:
$$g(z')=\phi_0+\sum_{i=1}^M \phi_i z_i'$$ where $M$ is the number of input features, $z'\in \{0,1\}^M$ are the features being observed or unobserved, respectively, and $z_i'=1$ or $z_i'=0$  and $\phi_i \in \mathbb{R}$ are the feature attribution values.\par
Given a model prediction $f(x)$, by assigning a feature mapping function $h_x(z')$ that maps binary inputs to the original feature space such that $x=h_x(z')$, we can evaluate $f(h_x(z'))$ and calculate the effect of observing or not observing a feature and seek to enforce $f(x) = f(h_x(z')) \approx g(z')$ through a special selection of $\phi_i$. This is one obvious desirable property requiring that the explanation model output matches the original model output. Shapley, through his work, described other desirable properties constraining the space of solutions for $\phi_i$: 
\begin{itemize}
  \item \textbf{Local accuracy/additivity/efficiency.} The sum of feature attributes need to match the original model output.
  \item \textbf{Missingness.} If a feature is missing, it receives zero attribution.
  \item \textbf{Consistency/monotonicity.} For two different models $f_1$ and $f_2$ in the same feature space, if the contribution of a feature $i$ increases for $f_2$ vs. $f_1$, then the given attribution for feature $i$ should not decrease for $f_2$.
  \item \textbf{Symmetry.} If $i$ and $j$ are two features that contribute equally,   their attribution should be equal.
  \item \textbf{Linearity.} The attributions of the sum of two functions $f_1$ and $f_2$ expands to the sum of the attributions for each of the two functions.
\end{itemize}

Those mathematically axiomatized properties (see \cite{lundberg_unified_2017} for details) describe a \emph{fairness} context of attribution. \par
Let $S \subseteq \mathcal{M}= \{1,\ldots,M\}$, a subset of non-zero indexes. By defining $f_x(S) = f(h_x(z')) = \mathbb{E}[f(x)|\boldsymbol{\operatorname{do}}(x_S)]$ then the only set of values \cite{shapley_value_1953}, \cite{young_monotonic_1985} for the explanation model satisfying the above properties can be proven to be:

\begin{equation}\label{eq:shap}
  \phi_i(f,x)=\sum_{S \subseteq \mathcal{M} \backslash \{i\}} \frac{|S|!(M-|S|-1)!}{M!}\Big[f_x(S \cup \{i\})-f_x(S)\Big]
\end{equation}

The above quantity represents some form of weighted average of the assigned attributions, calculated from model evaluation difference with and without the feature of interest, over all possible subsets of features $S$.\par

There has been some confusion in the literature over the proper evaluation function to be used to compute the feature contribution from the model \cite{aas_explaining_2019}, \cite{sundararajan_many_2019} \cite{janzing_feature_2019}. This confusion is due to ambiguity about which probability distribution the unconditioned variables should be averaged over, i.e., $\mathbb{E}[f(x)|x_S]$ vs. $\mathbb{E}_{x_{\bar{S}}}[f(x)]$. At first glance, as the minimizer of the squared loss, the former seems an appropriate and commonly used estimator to use, since the conditional expectation summarizes the whole probability distribution. However, several carefully designed counter-examples can be constructed (see \cite{sundararajan_many_2019}) to show that if the former conditional expectation is used as a basis for the functional calculation of Shapley values, then  $\phi_i \neq 0 \centernot\implies f \hbox{ depends on } x_i$, which violates the missingness property described earlier, i.e., the contraposition statement that if the feature $i$ is missing, it must receive an attribution of zero. In other words, if a feature exhibits no information with respect to the total outcome, it should not be influential. In this paper, we chose the computation of marginal expectation rather than the conditional; hence the presence of Pearl's $\boldsymbol{\operatorname{do}}$ operator from causal inference calculus \cite{pearl_causality_2000}. Therefore the remaining variables $x_{\bar{S}}$ are left untouched and are sampled from their natural distribution with no conditioning, as follows: 
$$\mathbb{E}[f(x)|\boldsymbol{\operatorname{do}}(x_S)]=\mathbb{E}_{x_{\bar{S}}}[f(x)]=\int \mathbb{E}[f(x)|x_S,x_{\bar{S}}] d\mathbb{P}(x_{\bar{S}})$$
where we note that by \cite{pearl_causality_2000}: $$\mathbb{P}[f(x)|\boldsymbol{\operatorname{do}}(x_S)]=\int \mathbb{P}[f(x)|x_S,x_{\bar{S}}] d\mathbb{P}(x_{\bar{S}})$$ denotes the distribution of $f(x)$ under intervention on $X_s=x_s$.\par
\begin{figure}[!t]
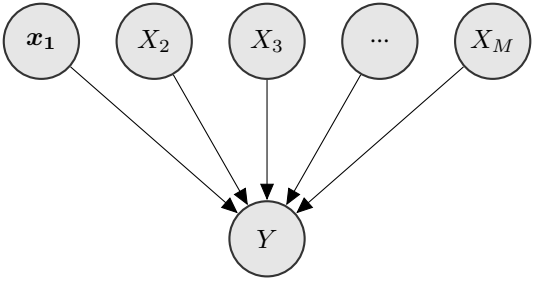

  \centering
  \tikz{
  \tikzstyle{main}=[circle, minimum size = 10mm, thick, draw =black!80,node distance = 16mm]
% nodes
 \node[main,fill = black!10] (Y) {$Y$};%
 \node[main,above=of Y,xshift=-3cm,fill = black!10] (X1) {$\boldsymbol{x_1}$}; %
 \node[main,above=of Y,xshift=-1.5cm,fill = black!10] (X2) {$X_2$}; %
 \node[main,above=of Y,xshift=0cm,fill = black!10] (X3) {$X_3$}; %
 \node[main,above=of Y,xshift=1.5cm,fill = black!10] (X4) {...}; %
 \node[main,above=of Y,xshift=3cm,fill = black!10] (XM) {$X_M$}; %
% edges
 \edge {X1,X2,X3,X4,XM} {Y}}
 \caption{Example of how unaffected variables sampled from the joint distribution to assess the influence of $X_1=x_1$ are intervened upon according to the rules of causal inference $\mathbb{P}_{X_2,\ldots,X_M}$}
  \label{fig:sampling}
\end{figure}

For clarity, given the graph structure shown in figure \ref{fig:sampling}, to evaluate the influence of a feature $X_1$ on the output $f(x)$ after observing $X_1=x_1$, we sample from the joint distribution of the remaining feature variables $\mathbb{P}_{X_2,\ldots,X_M}$.

It is clear from the expression (\ref{eq:shap}) that there are too many $\mathcal{O}(2^M)$ terms to evaluate the summation completely. As the number of features $M$ increases, the number of possible subsets increases exponentially. The computation efficiency is critical for feasible and timely generation of attribution values. Lundberg \cite{lundberg_consistent_2019} managed to apply a series of tricks and derive an algorithm for tree ensemble structures that reduces the complexity of exact computation of Shapley values from $\mathcal{O}(TL2^M)$ to $\mathcal{O}(TLD^2)$, where $T$ is the number of trees considered in the ensemble model, $L$ is the number of leaves, and $D$ is the depth of the trees.\par

\begin{figure*}[!t]
    \centering
    \includegraphics[width=\textwidth]{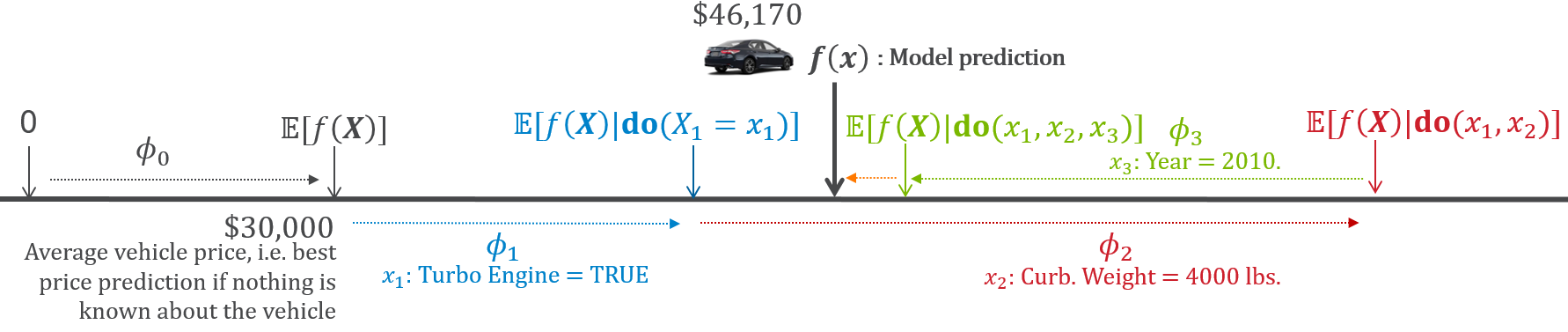}
    \caption{A single path walk-through of how attribution values are retrieved from successive model inquiries to sum to the model output prediction $f(x)$. The Shapley solution ensures that attributions are computed from an averaging over $M!$ possible orderings so that component boosting effects and interactions are taken into account.}
    \label{fig:shapwalk}
\end{figure*}

Figure \ref{fig:shapwalk} shows a single-path walk-through of how attribution values are retrieved from successive model inquiries. Shapley ensures fair $\phi_i$ values $\forall i$ by considering all possible combinations of sequences and orders. Shapley values are computed in relation to the reference baseline $\mathbb{E}[f(x)]$, assigned the attribution $\phi_0$. Therefore, the sum of the remaining attributions $i=\{1,\ldots,M\}$ captures the difference between the baseline value and the prediction. In our particular setting, Shapley values represent the change in price a certain component causes from the reference average price value of all the vehicles in the market (assuming the database is exhaustive). Although this is certainly valuable information, as we will show later, the true value of this exercise will emerge when selected vehicles are compared with one another, and component prices can be recovered. Also, attribution values can be aggregated over all vehicles with a focus on one component at a time, so component dependencies and their relationship to price will be better understood.

\begin{figure}[!t]
    \centering
    \includegraphics[width=2.5in]{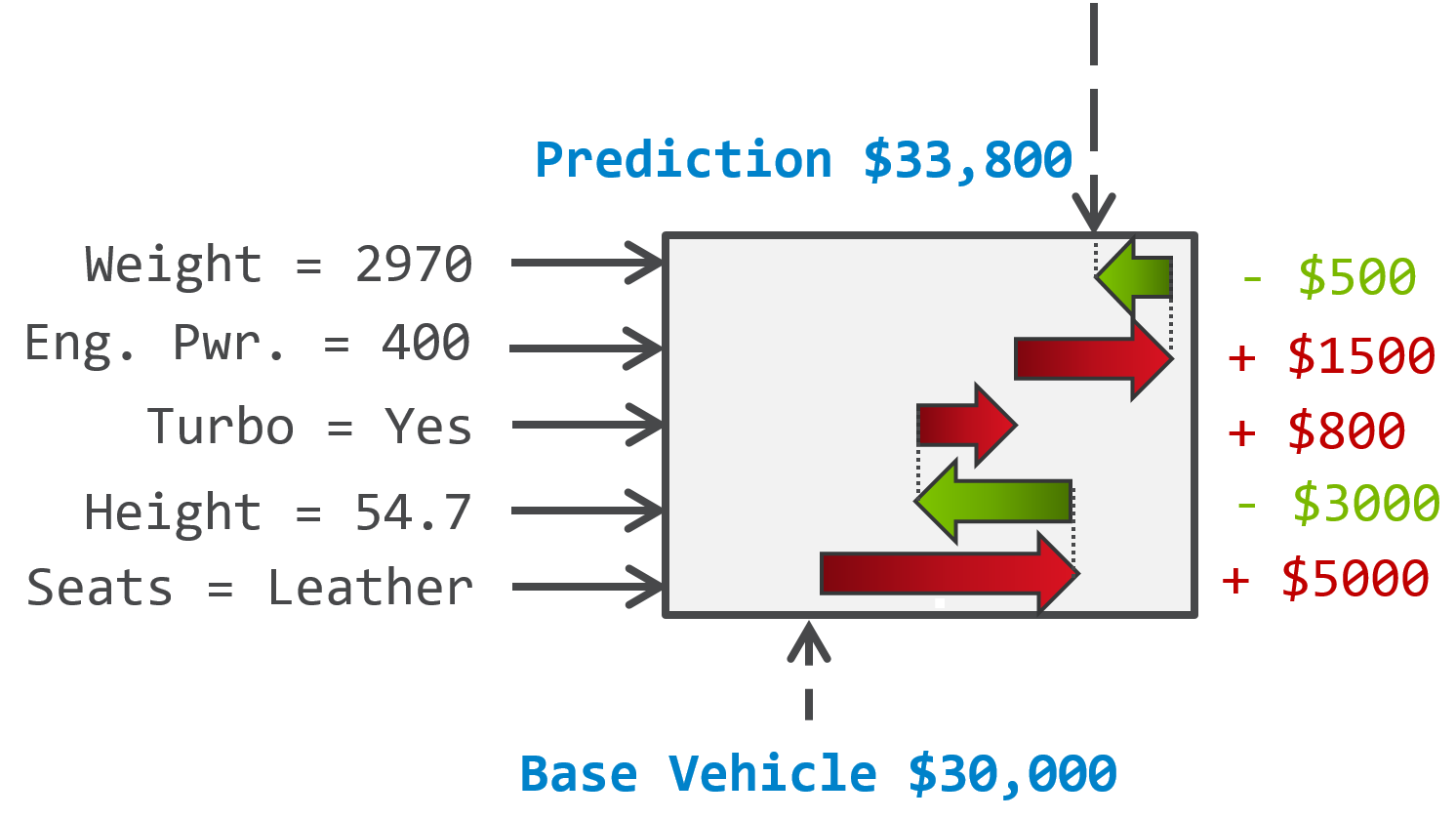}
    \caption{Example of how vehicle output can be broken down by giving each component a contribution to the outcome}
    \label{fig:shapbox}
\end{figure}

This theoretically supported and fair feature attribution method gives us the ability to better understand the contribution of each component to the vehicle's price. The next section presents a series of analyses at the vehicle level, where a single vehicle output can be broken down by giving each component a contribution to the outcome (figure \ref{fig:shapbox}), and shows how aggregation is achieved to get vehicle component prices from the total MSRP.

\subsection{Results}
\begin{figure}[!t]
    \centering
    \includegraphics[width=3in]{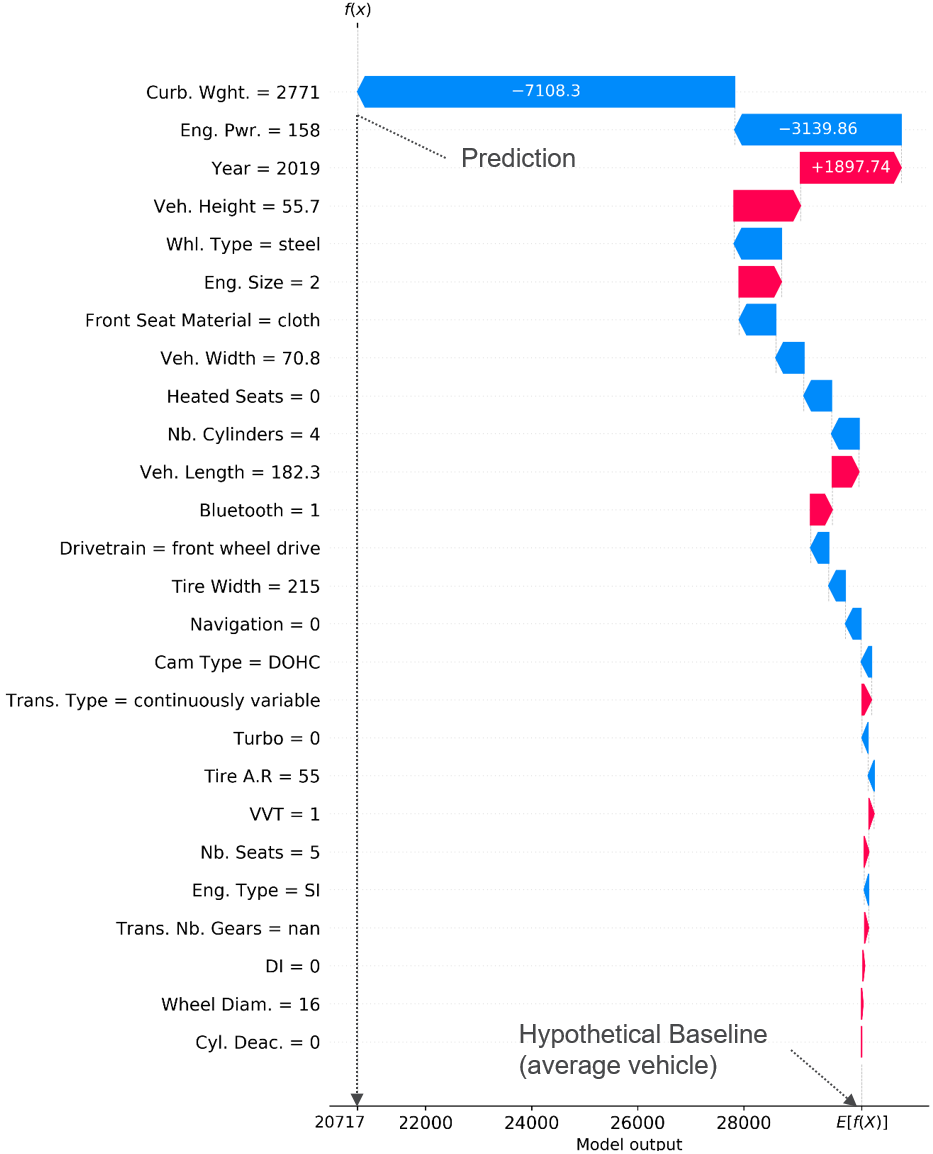}
    \caption{Example of vehicle MSRP prediction with the contribution values of individual components towards the pricing for the 2019 Honda Civic LX 4dr Sedan 2.0L 4 cyl CVT. The True MSRP is \$20,350, the predicted value is \$20,717.}
    \label{fig:oneveh}
\end{figure}

In this section, we show a series of component price estimation examples leveraging the Shapley attribution method.
At the unique vehicle level, figure \ref{fig:oneveh} presents an example of vehicle MSRP prediction with the contribution values of individual components towards the price. As explained earlier, Shapley values are computed in relation to a reference baseline vehicle represented by the market average vehicle, although this hypothetical vehicle is not necessarily useful in itself.

\begin{figure}[!t]
    \centering
    \includegraphics[width=3in]{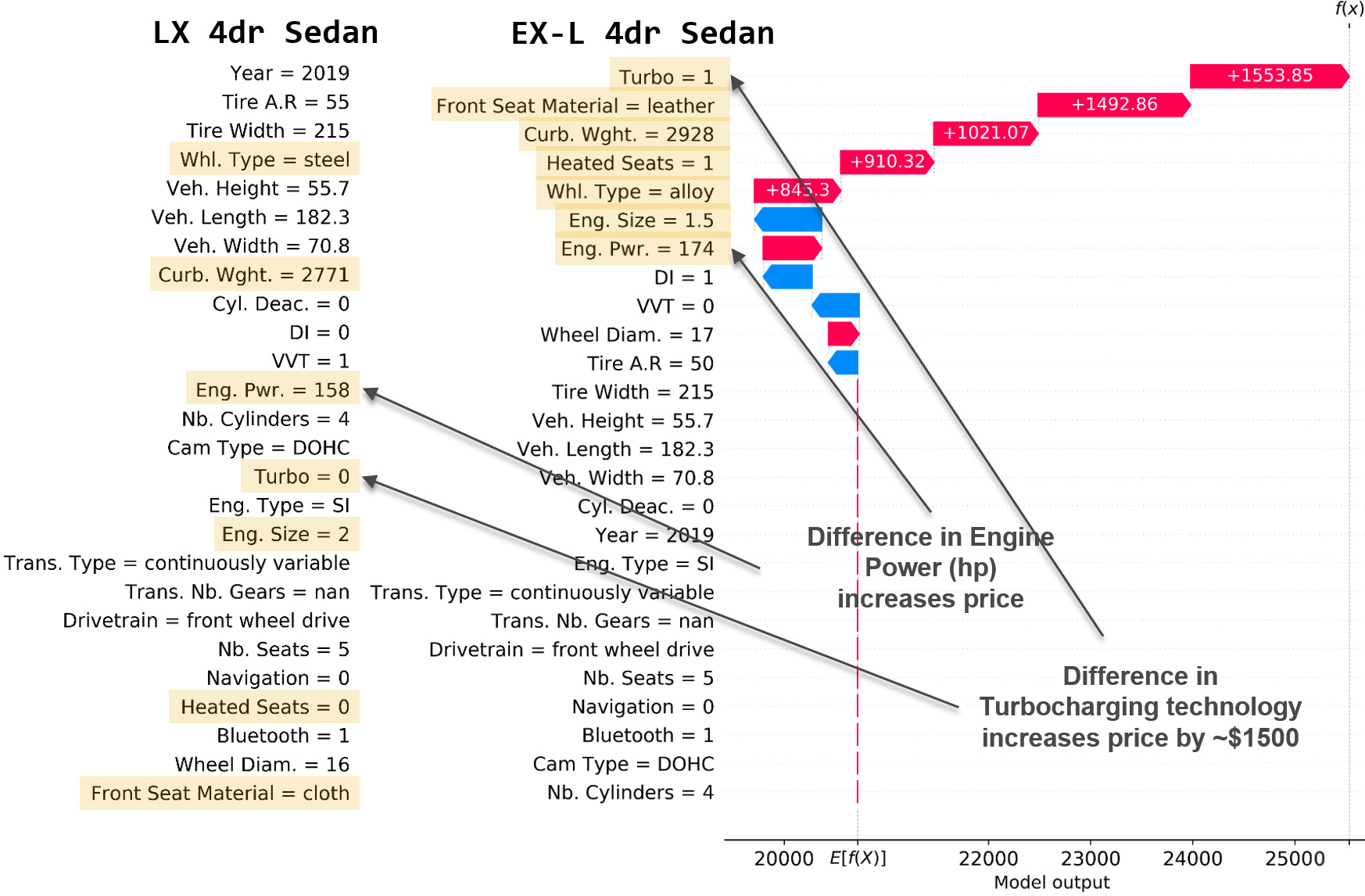}
    \caption{Example of trim level comparison and price difference explanation at the component level for two 2019 Honda Civic trims levels. The True MSRP for the EX-L 4dr Sedan trim is \$24,700, the predicted value is \$25,368.}
    \label{fig:twoveh}
\end{figure}

Figure \ref{fig:twoveh} shows a comparison of two trim levels of the same make and model vehicle for the same year (2019 Honda Civic). This direct trim level comparison allows us to better understand and quantify the components involved in the price difference. In this particular example, the advanced trim has advanced features that explain the price difference, and the additional price by component is computed through Shapley attribution. The turbo engine technology in the advanced trim vehicle explains an additional $\approx \$ 1500$ compared to the base trim level with no turbo technology, the alloy wheels contribute $\approx \$ 850$ compared to steel wheels in the base trim level, and so on.

\begin{figure}[!t]
\centering
\subfloat[Comparison of all trim levels of 2019 Honda Civic predictions. ]{\includegraphics[width=3in]{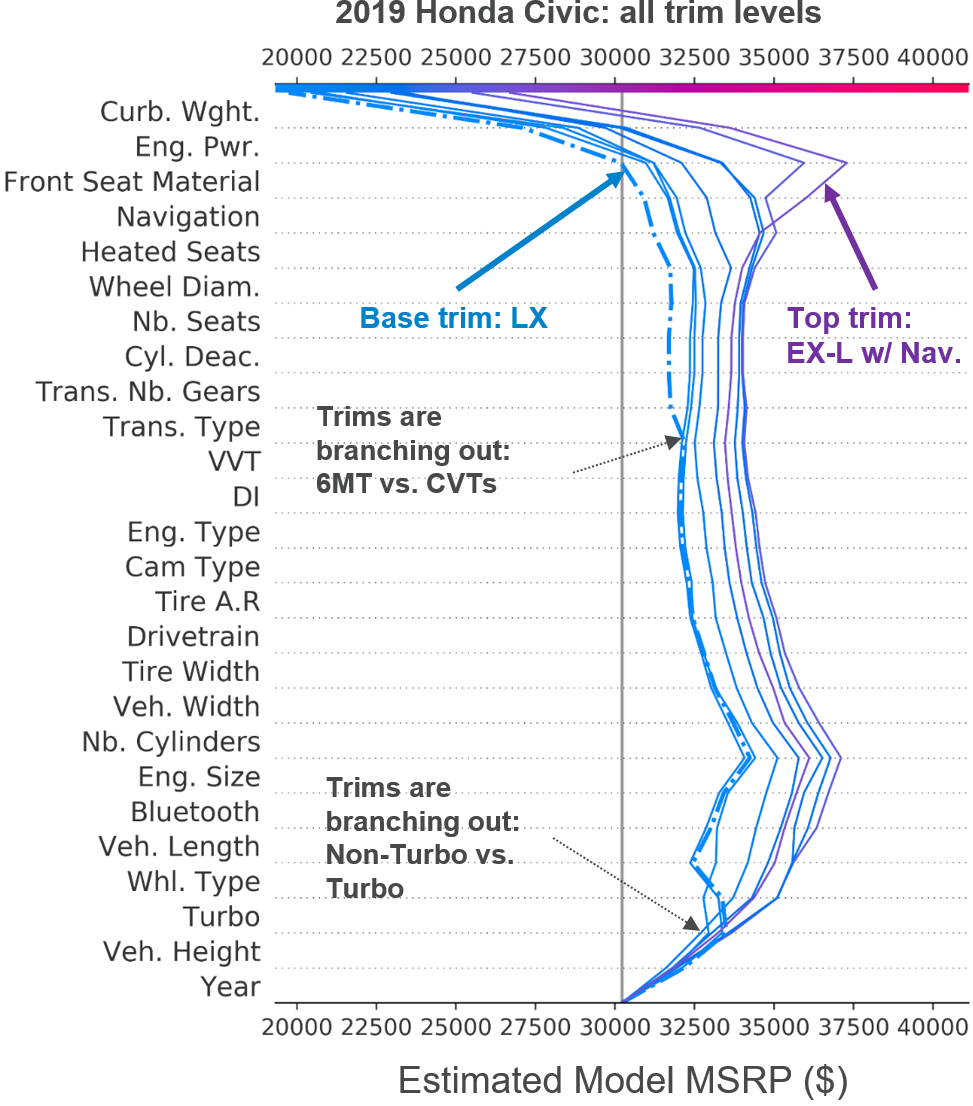}%
\label{fig:manyveh1}}
\\
\subfloat[Comparison of all trim levels of 2019 Toyota Highlander predictions.]{\includegraphics[width=3in]{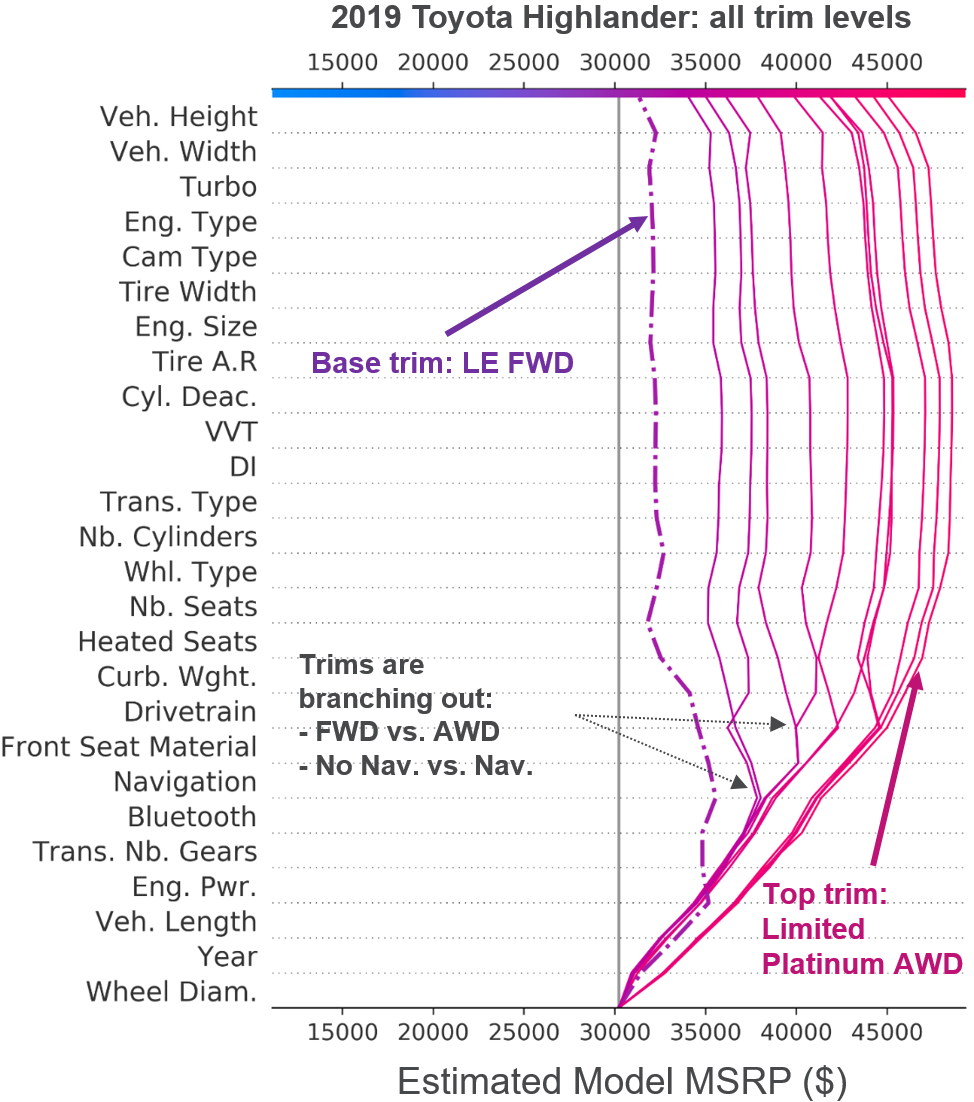}%
\label{fig:manyveh2}}
\caption{Example of several trim level comparisons and how component technologies affect the prediction path.}
\label{fig:manyveh}
\end{figure}

Increasing the number of vehicles in the comparison allows us to better understand the effect of some key vehicle components on pricing. Figure \ref{fig:manyveh} shows several trim levels of the same vehicle for two classes of vehicle (compact and SUV). The set of Honda Civic vehicles represents a typical compact class vehicle, while the Toyota Highlander represents a typical SUV class vehicle. We first note on the Civic graph that trim levels branch out in price with the inclusion of certain technologies. For example, the base trim level is the only trim that has a 6-speed manual transmission, while others have  continuously variable transmission (CVT) technology. This is clearly presented in the graph at the level of the transmission type, where all the slopes are  parallel except for one that indicates a decrease in pricing. On the other side, SUV trims seem to branch out for different reasons: The drivetrain type (all-wheel drive, 4-wheel drive, etc.) has a big impact on price. The low engine power of the base trim level seems to significantly decrease the price. It is worth noting as well that component technologies do not have the same effect on the two classes presented. For example, the vehicle height has a positive price impact on compact class vehicles (represented here by the Honda Civic) while the SUVs (represented here by the Toyota Highlander) show the reverse. These effects are in comparison to the hypothetical reference and clearly depend on the value of the component feature.

\begin{figure*}[!t]
    \centering
    \includegraphics[width=\textwidth]{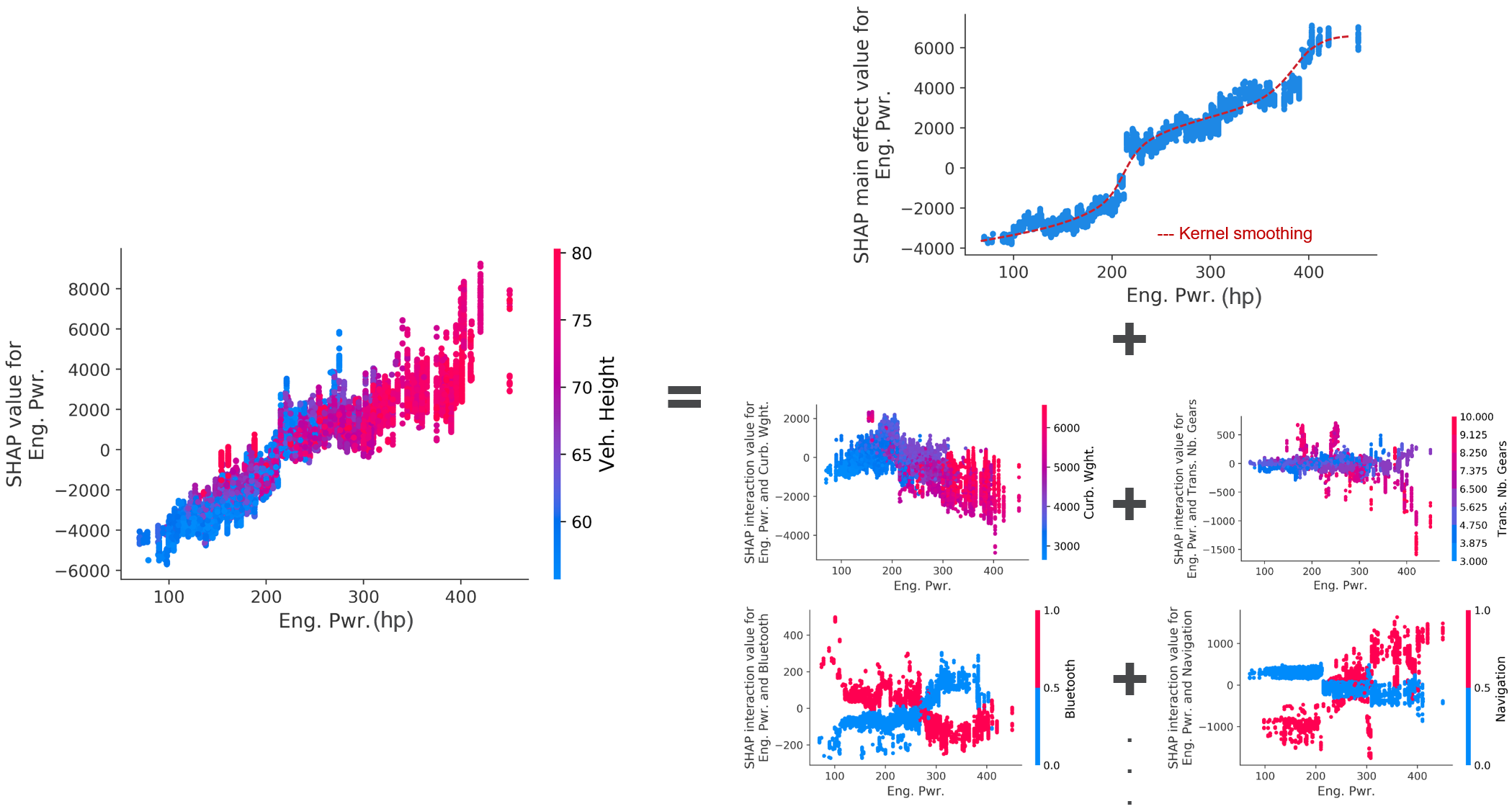}
    \caption{Dependency plot of the relationship between engine power attributed pricing and the feature value (left). The main/marginal effect (top right) removes the second order interactional effects. Interactional effects (bottom right) provide information about how vertical separation occurs due to strong interactions; here, engine power price attribution interacts quite strongly with the presence of technology options in the vehicle (Bluetooth, navigation systems).}
    \label{fig:shapagg}
\end{figure*}

Through the computation of Shapley attribution values for all the vehicles, and because every vehicle will have an distinct attribution value for each of its components, we can aggregate all the vehicles and focus on one component at a time and understand on a global level the overall effects that components have on prices. In figure \ref{fig:shapagg}, we show how individual component dependency plots can extract component price relationships by looking at the attributed Shapley value against the value of the feature of interest. This relationship shows how a feature attribution changes as the feature value varies. The left plot shows engine power dependency as the sum of the marginal effect, and the right plots all the second order interactions that engine power has with other components.\par
Retrieving the interactional effect gives valuable additional insights. We first recognize the complexity involved in component pricing with and without the presence of other specific components. Pricing is clearly performed in a "packaged" way, and this approach allows us to reverse-engineer the pricing strategies involved in this exercise. The variance displayed in the dependence plot for a given vertical slice is explained by the complex levels of interactions. To be clear, for example, a turbo system may be given a different price tag for a minivan than for performance car, or the price of a navigation system may be different if bundled with other advanced options than if purchased by itself. The marginal effect plot in the top right corner shows some vertical dispersion that accounts for beyond-second-order interactions.\par
Shapley interaction values are computed as follows:

$$\phi_{i,j} = \sum_{S \subseteq \mathcal{M} \backslash \{i,j\}} \frac{|S|!(M-|S|-2)!}{2(M-1)!} \nabla_{ij}(S)$$ where:
$\nabla_{ij}(S) = f_x(S \cup \{i,j\})-f_x(S \cup \{i\})-f_x(S \cup \{j\})+f_x(S)$ and the interaction is divided equally between feature $i$ and $j$ and $\phi_{i,j}=\phi_{j,i}$. The total interaction is given by $\phi_{i,j}+\phi_{j,i}$. The marginal effect can be extracted through $\phi_{i,i}=\phi_{i}-\sum_{j \neq i} \phi_{i,j}$ where we also note by additivity that $\sum_i \sum_j \phi_{i,j}= f(x)$. More details can be found in \cite{fujimoto_axiomatic_2006}

\begin{figure*}[!t]
\centering
\subfloat[Curb Weight (lbs)]{\includegraphics[width=0.33\textwidth]{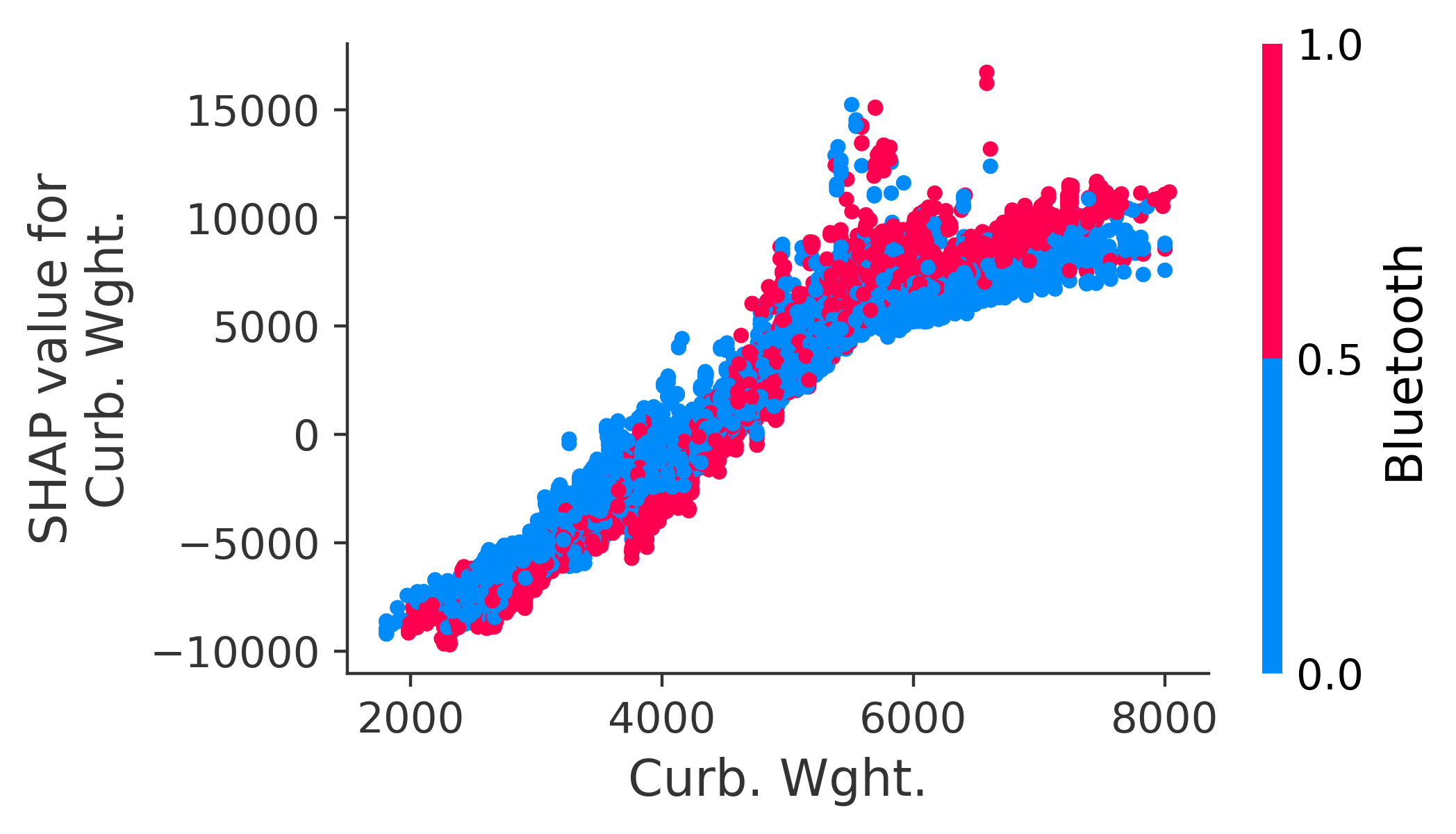}%
\label{fig:a}}
\hfil
\subfloat[Model Year]{\includegraphics[width=0.33\textwidth]{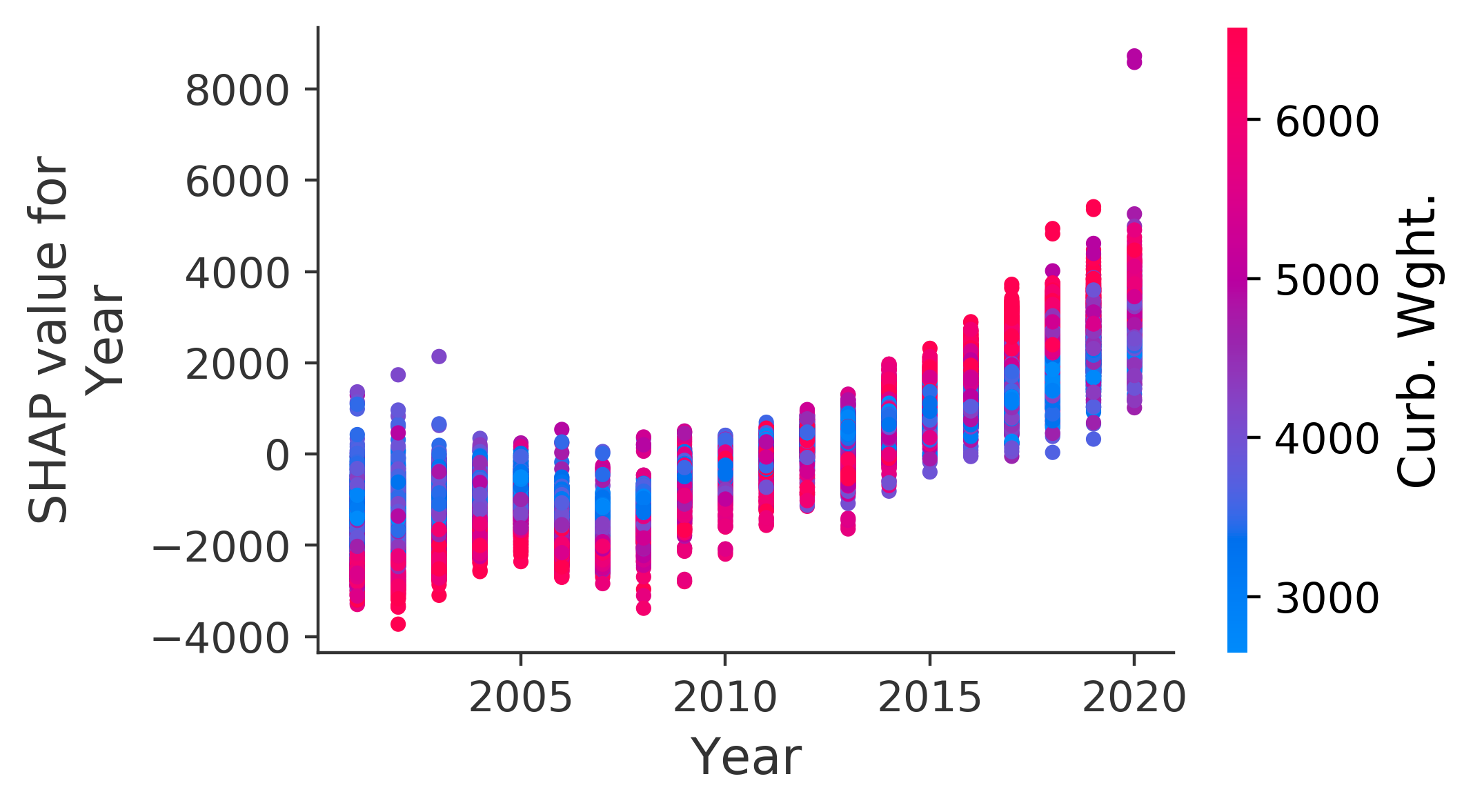}%
\label{fig:b}}
\hfil
\subfloat[Wheel Diameter (in.)]{\includegraphics[width=0.33\textwidth]{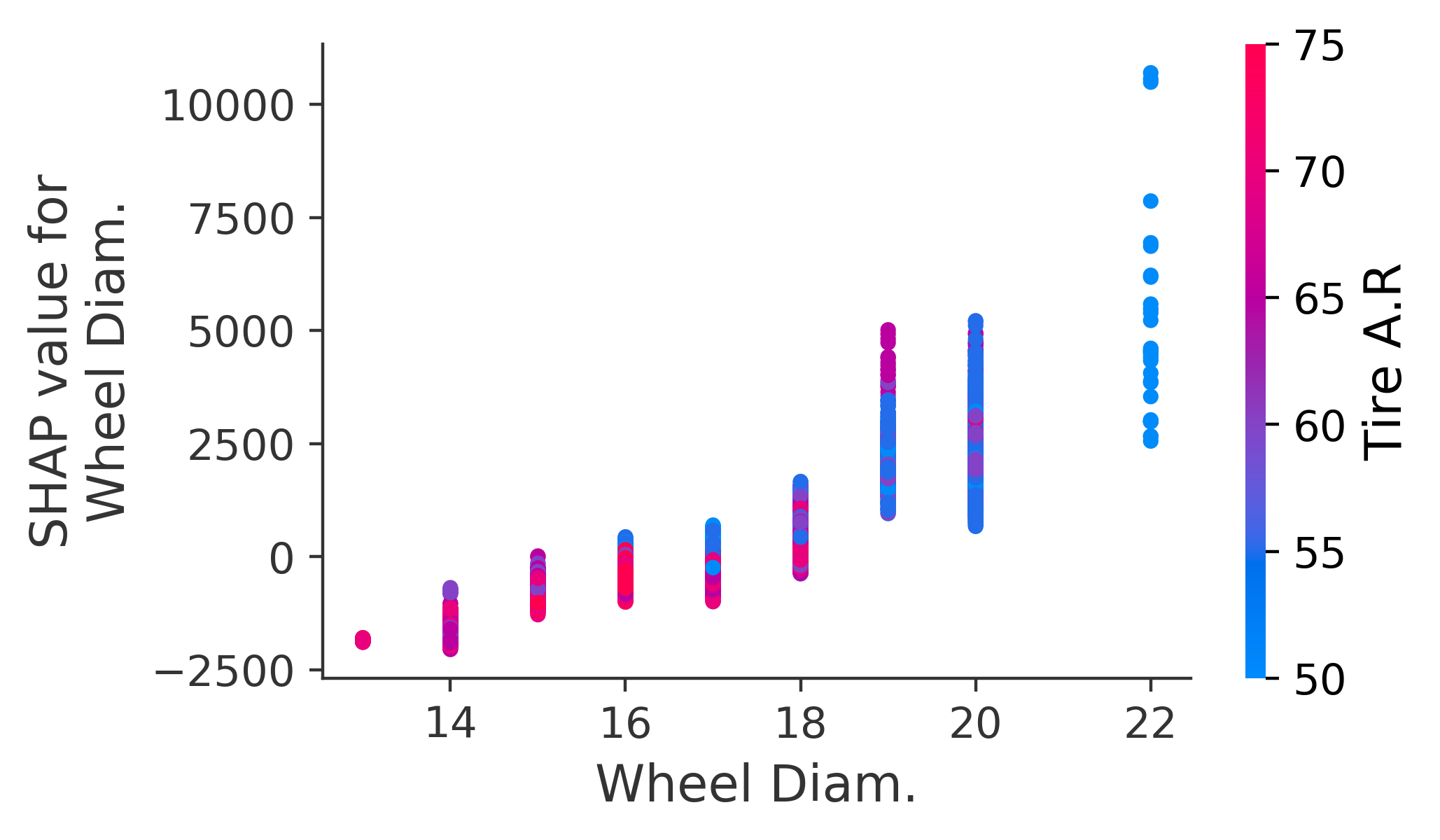}%
\label{fig:c}}
\\
\subfloat[Veh. Height (in.)]{\includegraphics[width=0.33\textwidth]{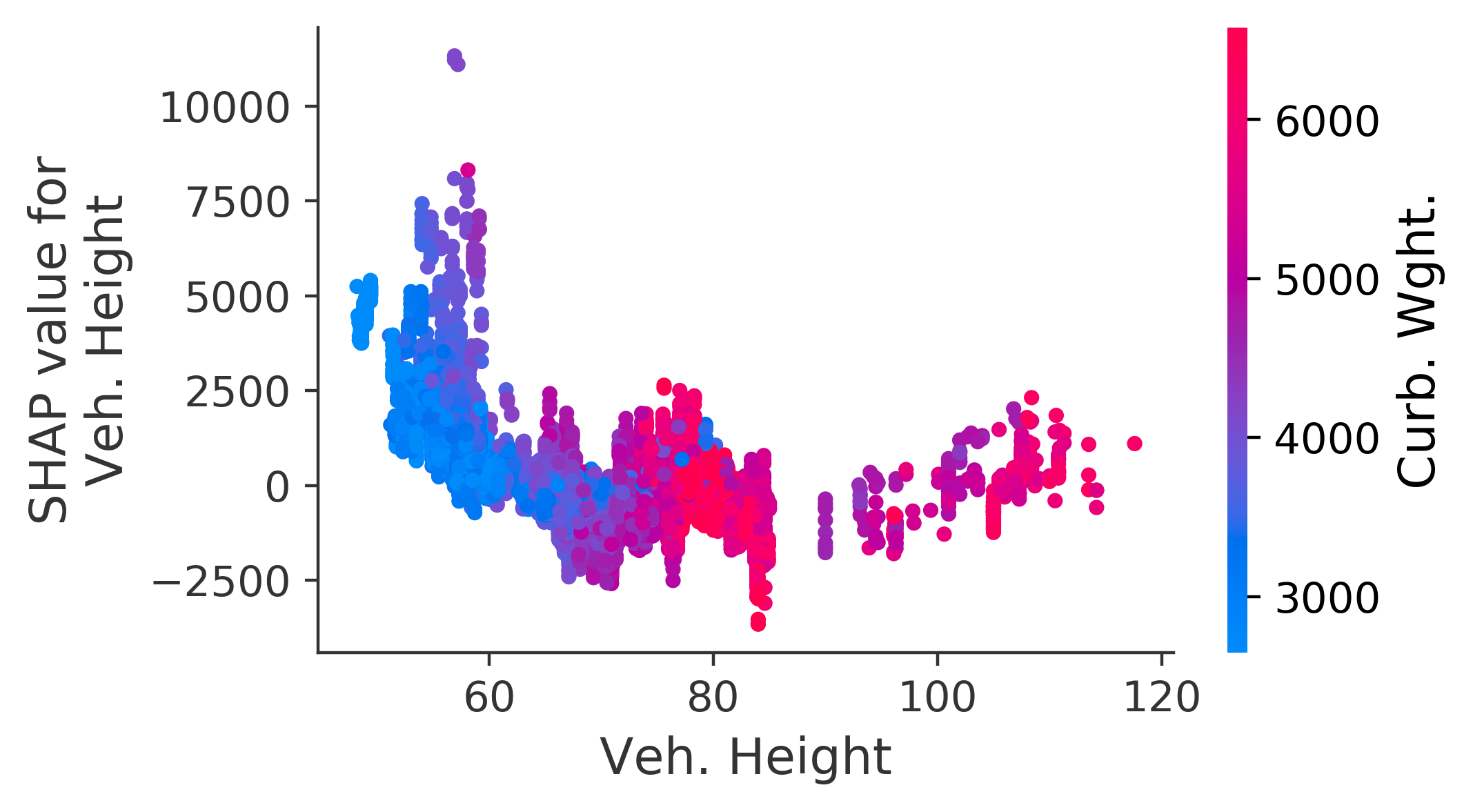}%
\label{fig:d}}
\hfil
\subfloat[Veh. Length (in.)]{\includegraphics[width=0.33\textwidth]{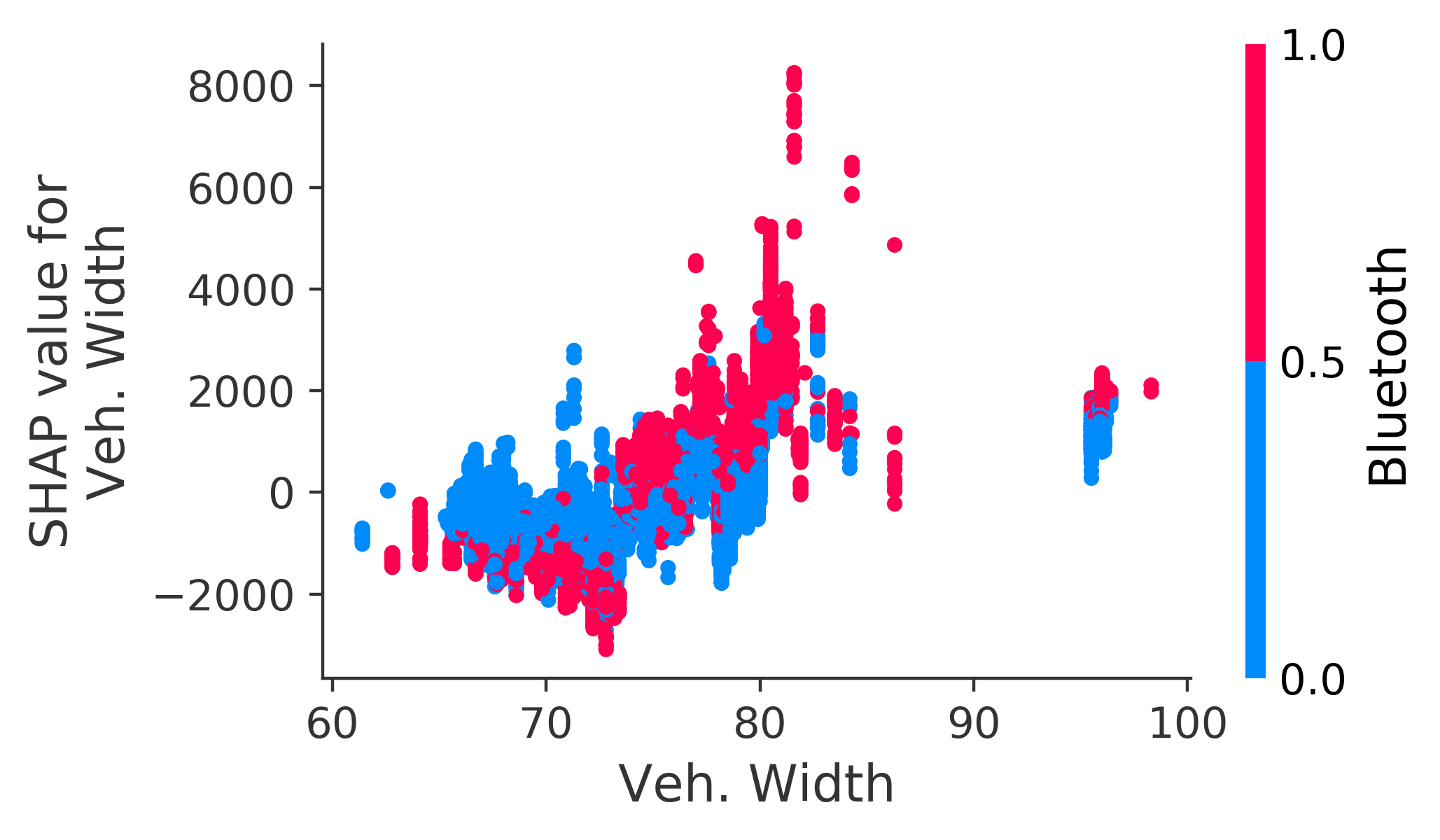}%
\label{fig:e}}
\hfil
\subfloat[Veh. Width (in.)]{\includegraphics[width=0.33\textwidth]{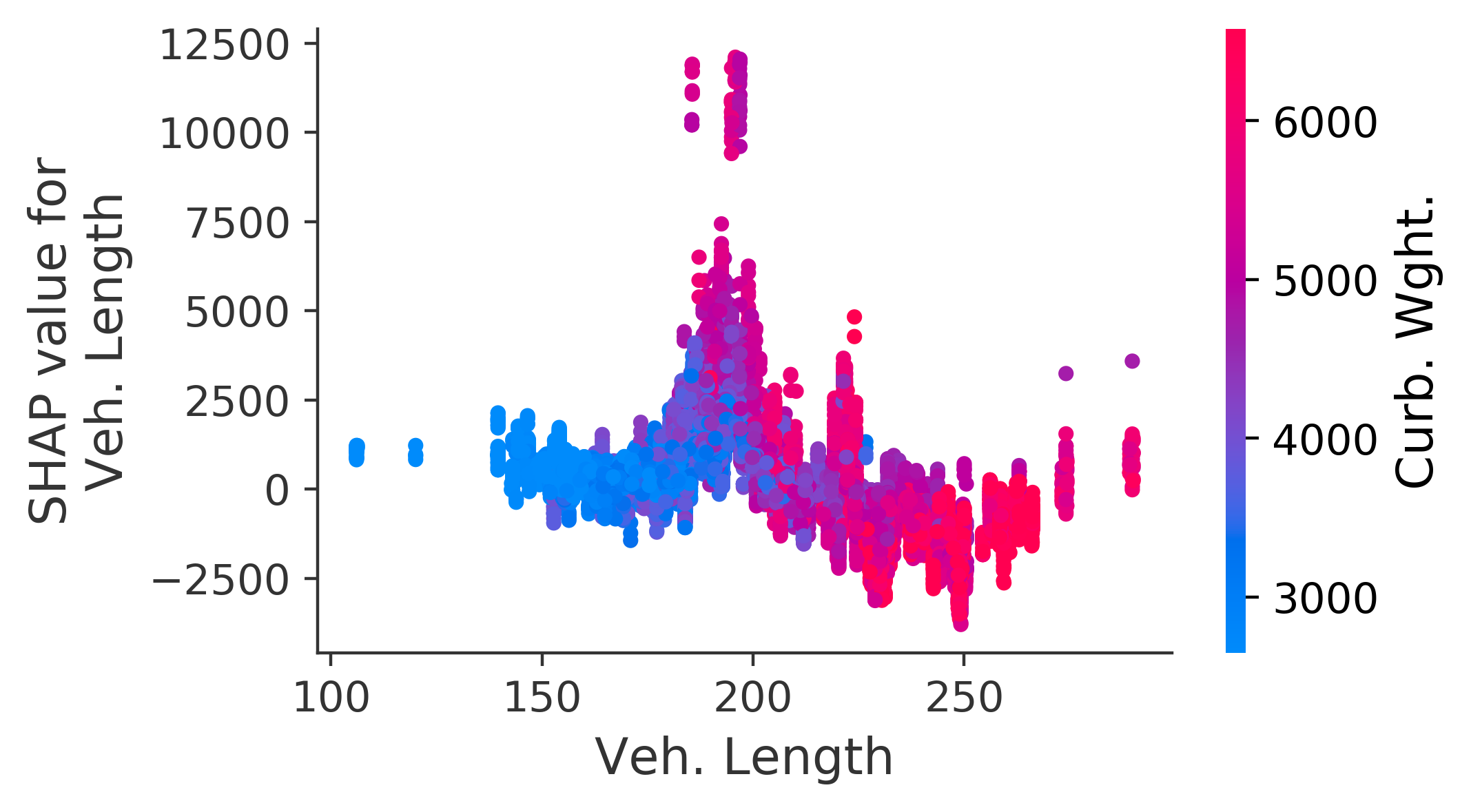}%
\label{fig:f}}
\caption{Dependency plots for a selected subset of key influential vehicle features.}
\label{fig:series}
\end{figure*}

We present the total effect dependencies of some top influential features in a series of plots in figure \ref{fig:series}. The marginal and interactional effect plots are omitted for a more concise analysis. We found strong non-linear dependencies (in vehicle curb weights in \ref{fig:a}, vehicle model year in \ref{fig:b}, vehicle height in \ref{fig:d}) and  quite complex dependencies (vehicle length \ref{fig:e}, and vehicle width \ref{fig:f}),\, while some components, like the effect of wheel diameter on price, could reasonably be approximated with a linear relationship (\ref{fig:c}). However, the presence of large vertical dispersion reveals the complex interactional effects involved in the pricing. For example, figure \ref{fig:b} shows the effect of the year on vehicle price, where we note a clear distinction in the trend between vehicles of larger curb weights vs. smaller curb weights. Heavier (ergo larger) vehicles seem to exhibit a sharper and more aggressive price increase over the years. In figure \ref{fig:a}, we see in the curves for the curb weight price relationship a separation between vehicles with and without Bluetooth. We underline that Bluetooth may not be the causal factor for this separation, as we explained in the causal graph in figure \ref{fig:sampling}, due to variable codependencies. The colors highlight the strongest computed interacting features.

We also observe other interesting trends and values for other components, such as the effect on pricing of the number of transmission gears, the type of wheels, tire characteristics, cylinder deactivation or other advanced engine technologies (direct injection, variable valve timing, variable valve lift, etc.), front seat material, etc. For a succinct paper, we omit them from the current analysis, and we plan to provide further analysis in a separate paper in the future.\par

\subsubsection*{Engine Displacement} 
\begin{figure}[!t]
\centering
\subfloat[Total effect: engine displacement (L). ]{\includegraphics[width=3in]{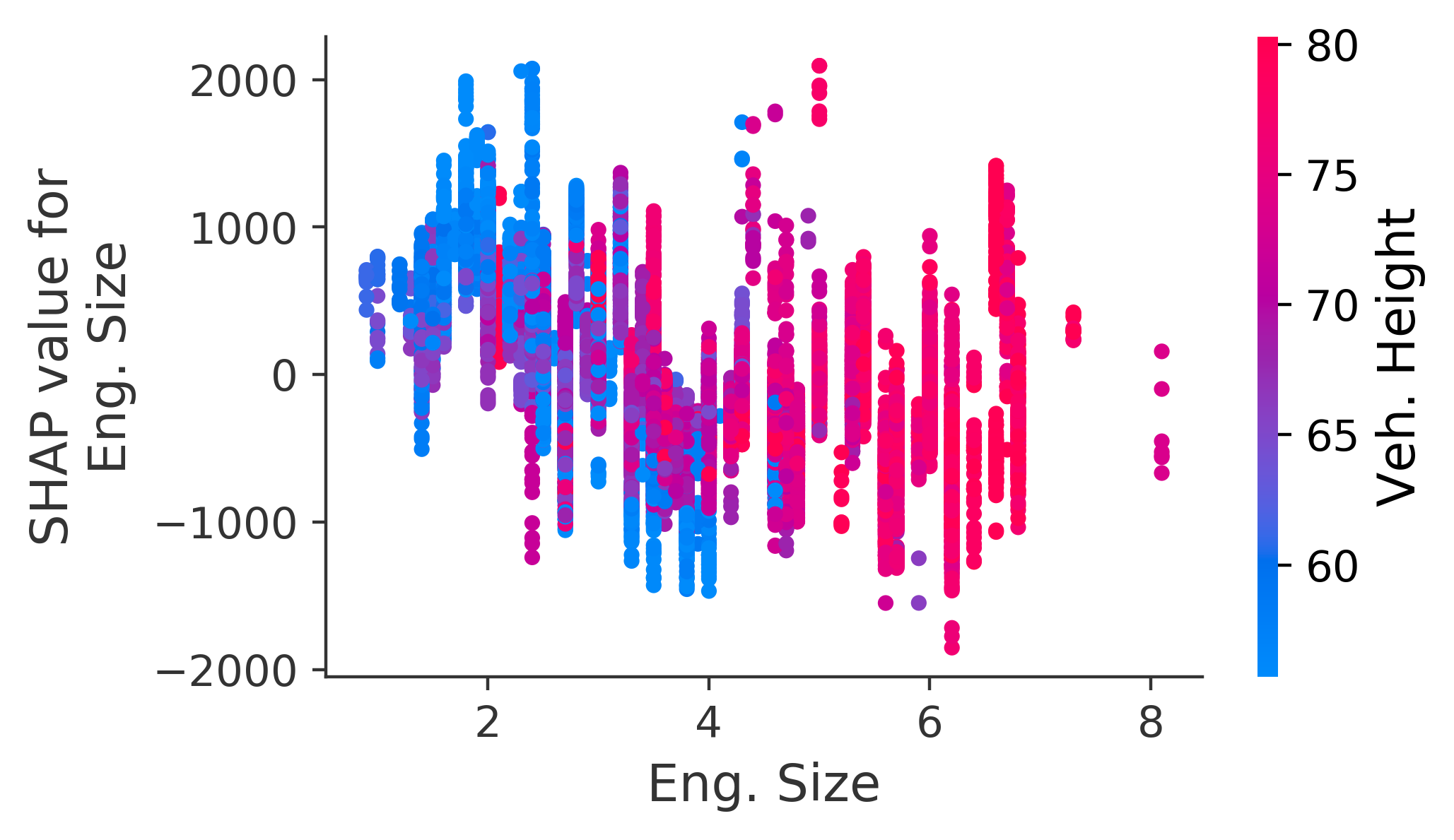}%
\label{fig:aa}}
\\
\subfloat[Main effect: engine displacement (L).]{\includegraphics[width=3in]{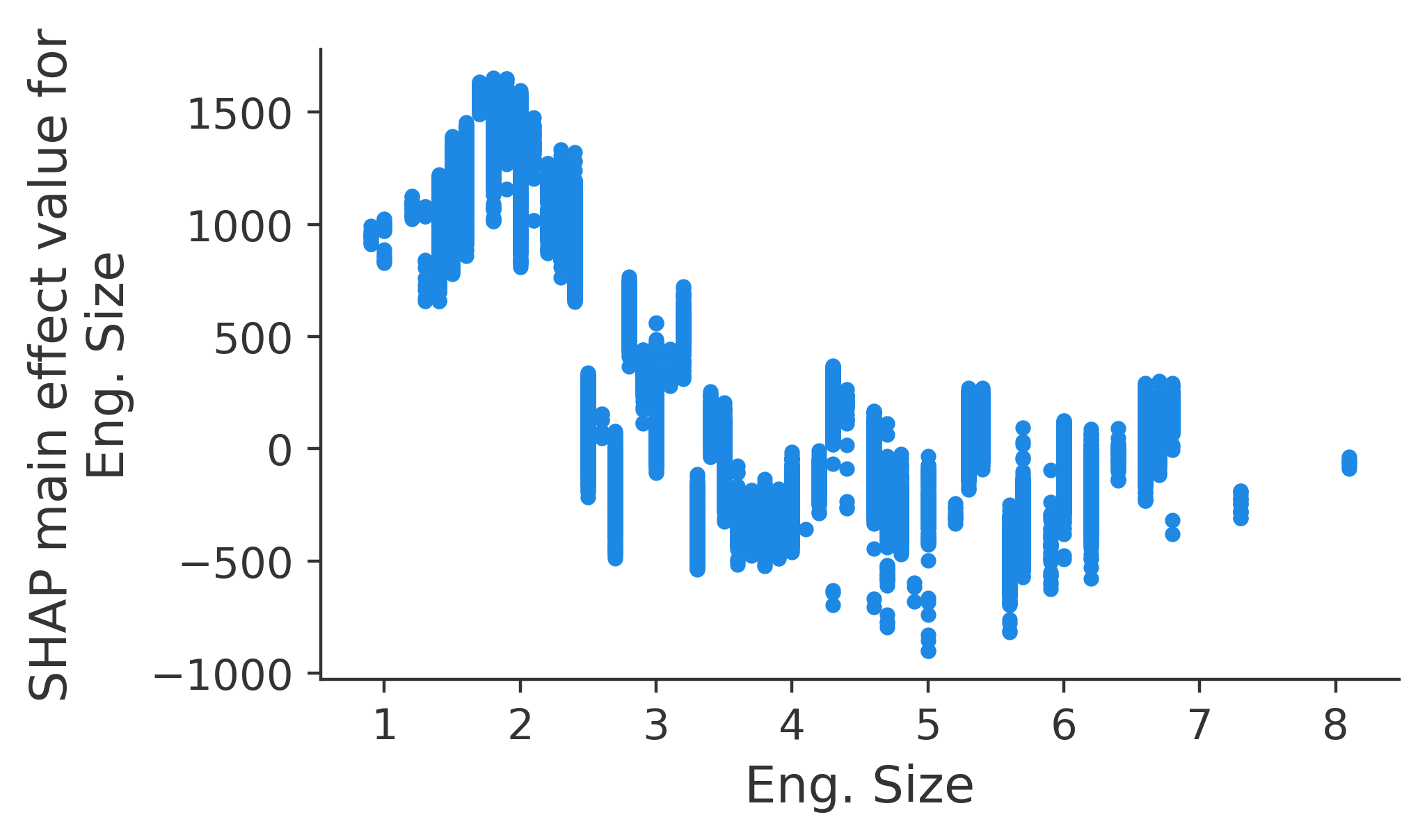}%
\label{fig:bb}}
\caption{Total effect (a) and main effect (b) of engine size (in liters) value on pricing}
\label{fig:engsize}
\end{figure}

Certain component price trends require special attention in their interpretation. In figure \ref{fig:engsize}, we consider the case of engine displacement's dependency on pricing. A remarkable, perhaps not so surprising, relationship emerges where we note an overall downward price trend with augmented engine size (displacement in liters).  A marginal increase in engine size has the effect of reducing price. By "marginal" we mean that other components are controlled for, and therefore the pure isolated effect of an increased engine size is contributing to the downward tendency. In figure \ref{fig:aa}, we show the total effect of the engine size feature's value on price, and the large vertical dispersion due to high levels of interactions obscures the small movements in the trend. In figure \ref{fig:bb}, the true effect is shown when second order interaction levels are removed for a clearer picture. We observe a more detailed change in direction at several key engine size levels, particularly at around 1.8-2 L, at 3.5-4 L and $\sim$ 5 L. From 1 L to 1.8-2  L engines, the tendency is upward; that is, an increase in engine size does contribute to an increase in the technology price. In this case of small cars, the higher price is possibly justified by the better fuel economy seen in these vehicles. However, starting from 1.8-2 L to 3.5-4 L we see a decrease in price.  The higher price of smaller engines in this portion of the graph can be explained by the fact that the reduction in engine size is made possible by better engine technology. Turbocharged 4-cylinder engines have replaced 6-cylinder engines in many applications. While the displacement of the turbocharged engines is smaller than the naturally aspirated ones, the use of additional components for turbocharging will increase the price of engine. Assuming the power output needed from the engine remains the same, getting it from a smaller displacement requires technologies such as turbocharging or higher compression ratios, which are all likely to cost more than the base engine. The larger engines in that list are all likely to be older, naturally aspirated engines in minivans or pickups. In case of minivans, people may not pay for more power or displacement. Hence the wiggling of the pricing curve between 4 L and 6 L engines. For pickups, the turbocharged engines with a lower displacement and higher power can command a higher price than a larger naturally aspirated engine. 

\begin{figure}[!t]
    \centering
    \includegraphics[width=3.5in]{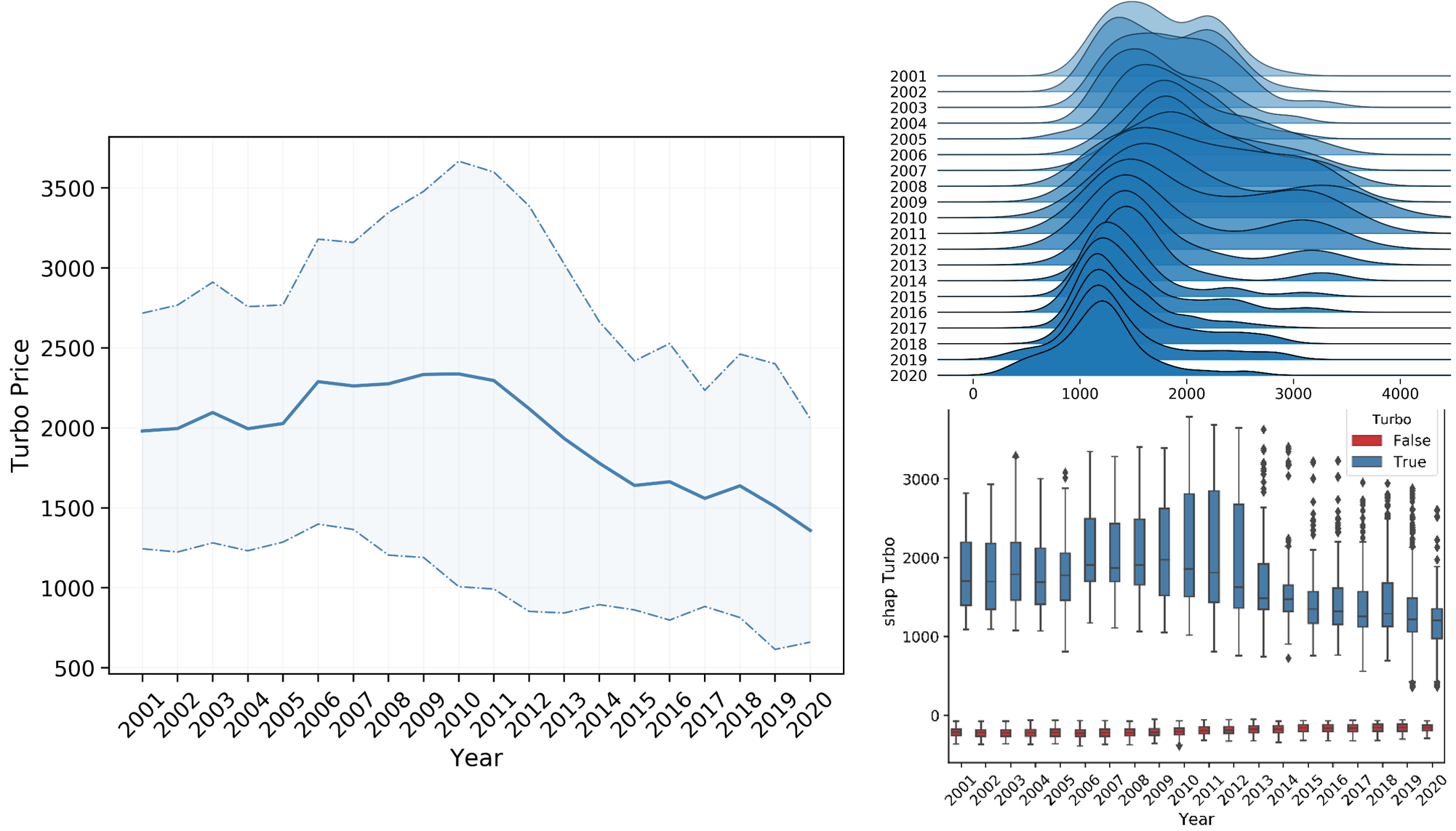}
    \caption{Effect of time on specific component pricing: turbo technology}
    \label{fig:turbotime}
\end{figure}

\begin{figure}[!t]
    \centering
    \includegraphics[width=3.5in]{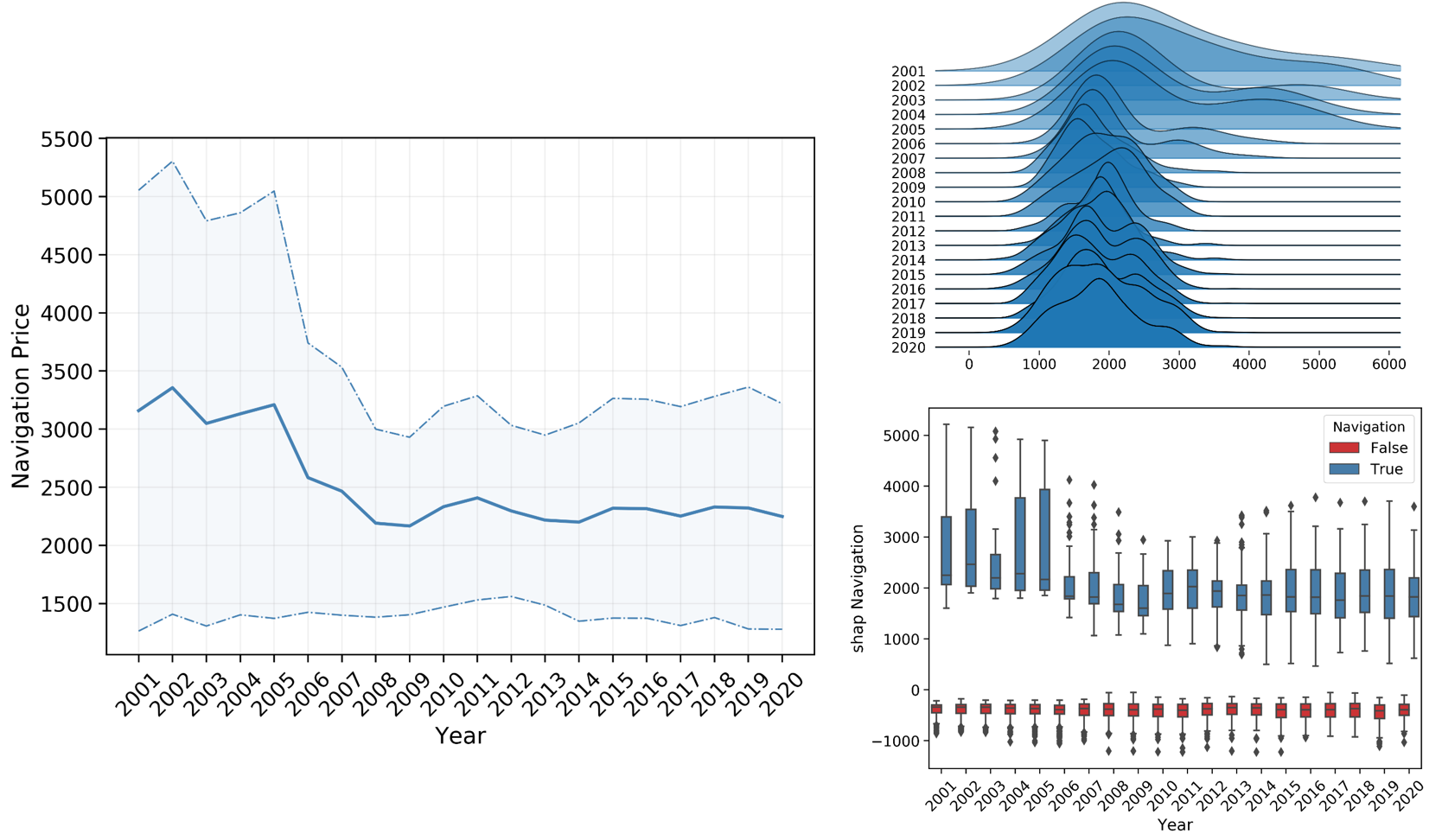}
    \caption{Effect of time on specific component pricing: navigation system}
    \label{fig:navtime}
\end{figure}

Individual technology prices can be assessed and studied over time. For example, figure \ref{fig:turbotime} shows the effect of time on turbo technology pricing, where after a period of slight increase (late 2000 economic recession), there appears to be a drop in turbo pricing. This is a clear example of how component prices evolve as they become more popular. Figure \ref{fig:navtime} gives another non-powertrain example of such behavior: navigation system price evolution. We observe here  fairly stable pricing since 2010, but perhaps a slight, steady, but almost indistinguishable decrease.
Analogous examples can presented to follow pricing changes over time when a certain technology (e.g., a rear camera) is made compulsory through regulation (not shown here).

\begin{figure}[!t]
    \centering
    \includegraphics[width=3.5in]{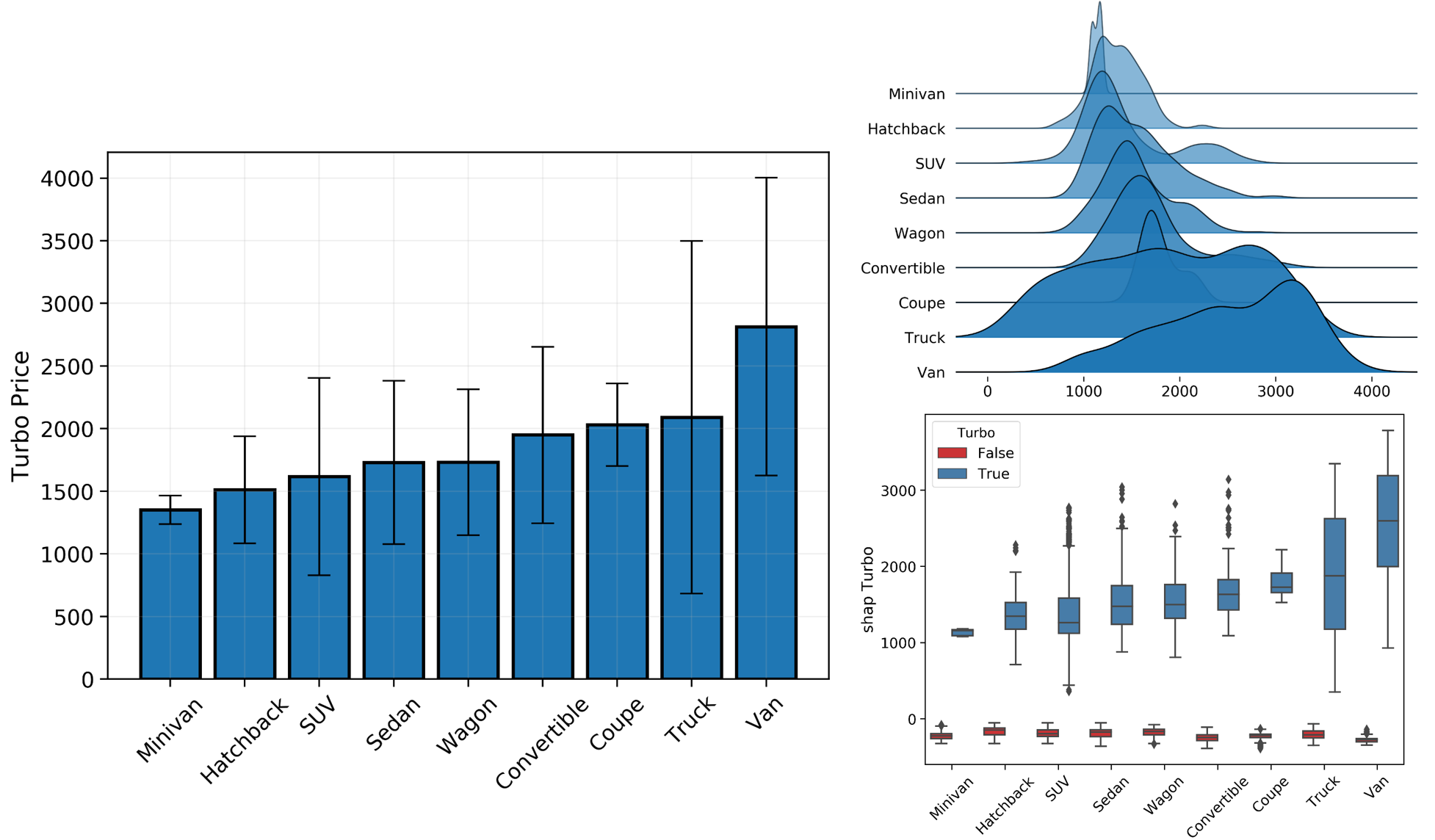}
    \caption{Effect of class on specific component pricing: turbo technology}
    \label{fig:turboclass}
\end{figure}

In a similar fashion, the effect of vehicle class on component prices can be assessed. Figure \ref{fig:turboclass} shows clear evidence of how turbo pricing is affected by the vehicle segment. We note an overpricing of the technology for trucks and vans, while minivans display the lowest price, manifestly due to the type of customer who is not necessary seeking efficiency and is not willing to pay for it. On the flip side, vans are typically used as the primary carrier for delivery options. In this category, the benefit of turbocharging is worthwhile and in high demand and is therefore priced differently.

\section{Influential Vehicle Features}
The Shapley values can be used to identify the importance of features to the model output (i.e., to the vehicle price prediction). Features have the most impact when the change in the model output is greatly affected by the feature value. For linear models $f(x)=x^T\beta$, the coefficients of the covariates provide some clues. In a typical setting, the importance of a feature is given in a global form; that is, the importance is measured by looking at its effect on model error. For example, the permutation feature importance method consists of measuring the model prediction error after permuting the values of a specific feature. Then the one feature that most affected the model accuracy is attributed a quantifiable high importance: a global measure of how the model reliability is dependent on that feature. Conversely, a feature is not important if the resulting permutation did not affect the model error (see \cite{statistics_random_2001} or \cite{fisher_all_2019} for an exploration of other  methods).\par

Alternatively, because of its natural local property, the attribution provided by Shapley derivation gives an individualized feature importance measure for each prediction and each feature. Their aggregation can ultimately provide an equivalent global importance measure, but the natural decomposition produces a richer view of importance. In fact, typical feature importance plots are  bar charts showing the general effect a feature has on the prediction, while the Shapley approach, endowed with localizable importance values, delivers higher resolution plots. In addition, Shapley solutions ensure consistency in the sense stated above in section \ref{sec:shapmeth}.

\begin{figure}[!t]
\centering
\subfloat[]{\includegraphics[width=3in]{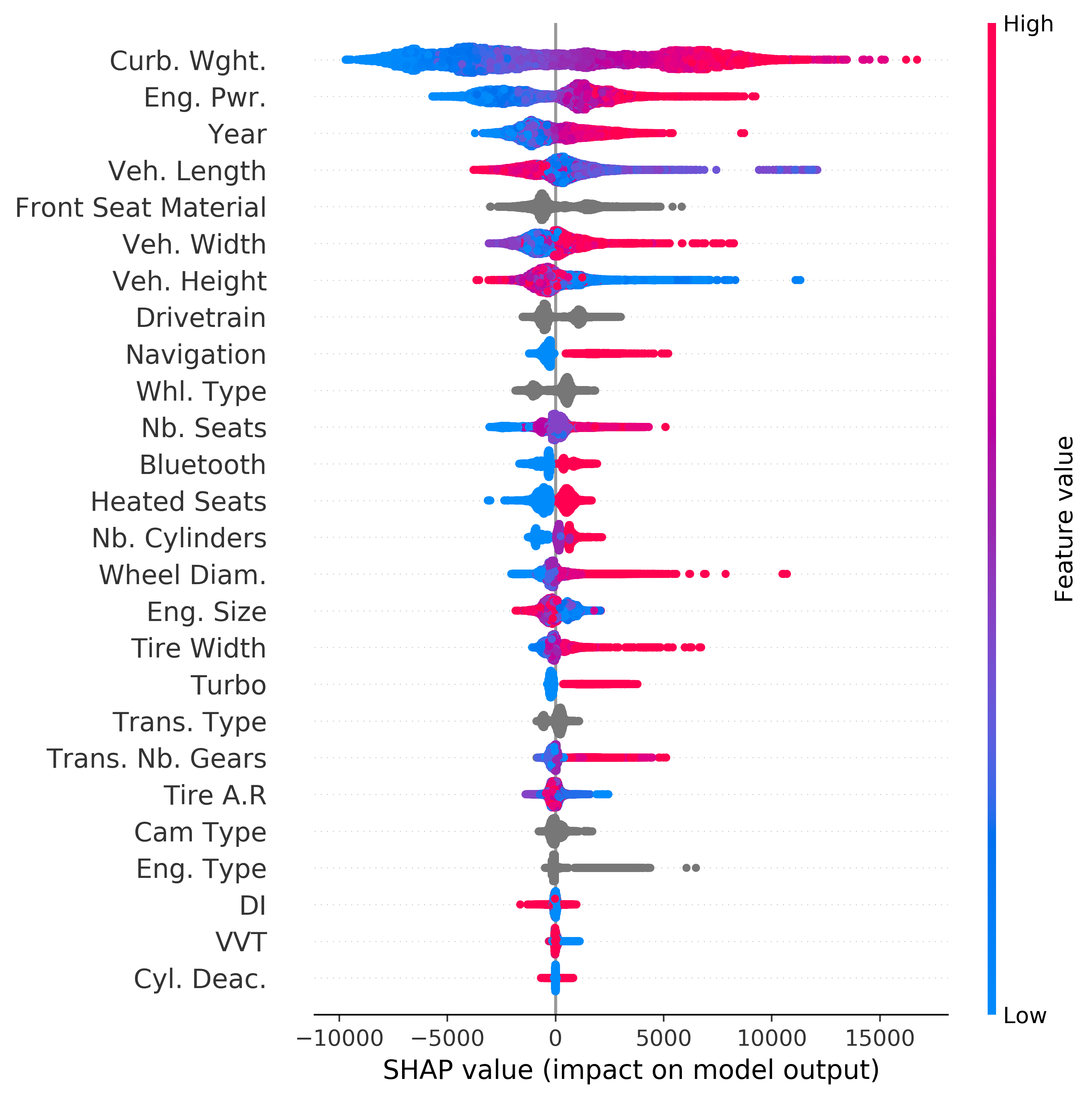}%
\label{fig:summary}}
\\
\subfloat[]{\includegraphics[width=3in]{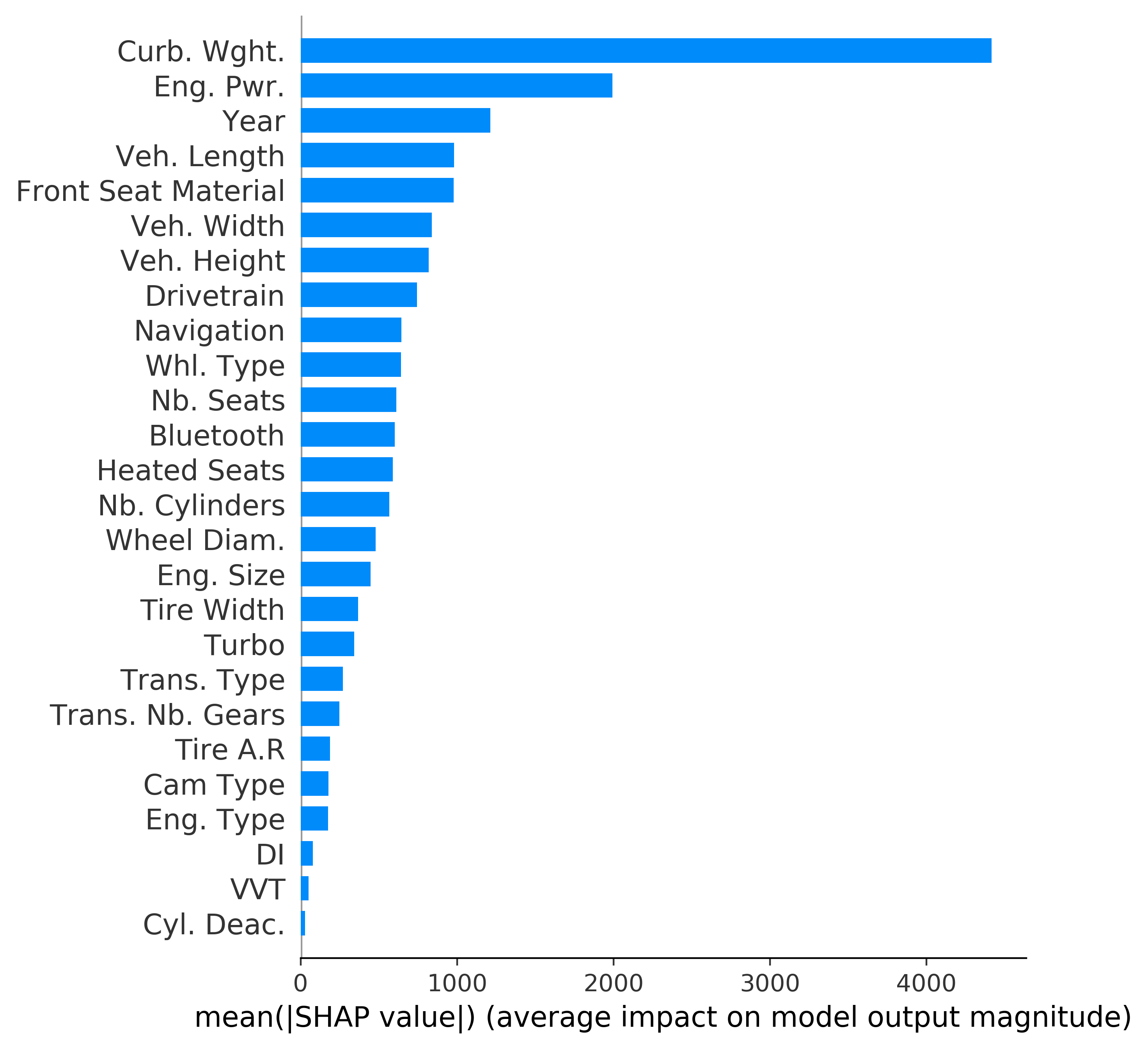}%
\label{fig:summarybar}}
\caption{(a) Individual (one dot per vehicle) Shapley attribution values for a subset of features of the Catboost MSRP predictive model. High Shapley values mean a high price attribution, which depends on the feature value shown by the color code. (b) Standard feature importance bar chart.}
\end{figure}

% \begin{figure}[!t]
%   \begin{subfigure}[b]{0.5\textwidth}
%     \includegraphics[width=\textwidth]{summary}
%     \caption{}
%     \label{fig:summary}
%   \end{subfigure}
%   \hfill
%   \begin{subfigure}[b]{0.5\textwidth}
%     \includegraphics[width=\textwidth]{summarybar}
%     \caption{}
%     \label{fig:summarybar}
%   \end{subfigure}
%   \caption{(a) Individual (one dot per vehicle) Shapley attribution values for a subset of features of the Catboost MSRP predictive model. High Shapley values mean a high price attribution, which depends on the feature value shown by the color code. (b) Standard feature importance bar chart.}
% \end{figure}

Figure \ref{fig:summary} shows the individual Shapley attribution values for a subset of features of the Catboost MSRP predictive model. High Shapley values mean a high price attribution, which depends on the feature value shown by the color code. The plot gives a high resolution feel for feature importance, as each dot is a vehicle feature attribution value. In  this type of plotting, the amplitude provides a general idea of the overall distribution over Shapley values that each feature has. The features are ordered by order of importance by summing over the $N$ vehicle examples $j$, i.e., $\frac{1}{N}\sum_{j=1}^N |\phi_i(f,x_j)|$ for each feature $i \in \mathcal{M}$. Figure \ref{fig:summarybar} shows the standard feature importance bar chart computed by the formula just described, which measures global importance through summation over all vehicles. For example, with the vehicle curb weight topping the list, the plot shows that the vehicle curb weight is the most influential variable affecting vehicle pricing. The higher the Shapley value, the bigger the contribution to the total price the curb weight has, and from the colors we see clearly how the higher vehicle curb weight feature value increases the pricing (unsurprisingly). The large variance also provides information on the spread of the vehicle curb weights in the dataset, and the density shows how common each is. We see roughly  five density groupings, most likely corresponding to the five main standard U.S. Environmental Protection Agency (EPA) vehicle classifications.\par

It is worth noting that no causal conclusion can be drawn from the above analysis, and that vehicle curb weight increase or decrease is not causing vehicle price to change. Vehicle curb weight as encoded in the current model should be interpreted as a proxy for other latent parent variables. Observing one of those variables will change the distribution over curb weight due to the dependencies and therefore its overall influence. In other words, this is a property of the built system and model, but not of the external world.

\section{Discussion}
\subsection{Methodology Implementation}
Current vehicle pricing methods rely on fixed equations to calculate each component cost or technological incremental cost and, ultimately, vehicle manufacturing cost. The vehicle's MSRP is then computed using a constant 1.5 multiplier for retail price equivalent (RPE). In reality, as we showed, OEMs have different margin levels based on the vehicle class, vehicle technology, or other criteria.\par
In this paper we proposed an alternative no-teardown top-down approach for component and vehicle price estimation, but improvement upon the proposed work and an extension of its application is possible. Promising new studies can build on the Argonne vehicle attribute database and the developed technique. In particular, further studies can be performed to estimate the increased efficiency per unit of price (\$/mpg) to increase the reliability of overall VTO benefits. This opens the door to deriving \$/mile estimates at the vehicle technology level and also deriving component-level \$/mile estimates to explore the tradeoffs between more efficient vehicle technologies (powertrain level or component) and the added price. Connecting those estimates with sales data will enable an understanding of technology's value to the customer.\par
This novel proposed methodology shows some advantage over current Autonomie vehicle pricing methods. Profiting from this novel methodology for future VTO-related benefits analysis efforts would require significant Autonomie process changes. There are two options for direct implementation of this novel methodology into the Autonomie framework:
\begin{enumerate}
  \item \textbf{Equation based.} Preserve the current Autonomie methodology and derive parametric equations (or non-parametric relationships, e.g., kernel smoother methods) for each component and implement independent component prices at the MSRP level (including direct and indirect costs). There will be no need for post-hoc RPE or ICM adjustment. However, due to the high degree of interactional effects, this approach is not recommended.
  \item \textbf{Shapley-based credit/penalty component pricing.} This approach would rely on the use of the predictive model to estimate vehicle price and then generate the Shapley values to extract a price contribution for each component. Starting from a baseline vehicle and component value, a price credit/penalty is applied via the Shapley attributed score: Through the complexity of interactional effects, the price of a component will be dependent on the presence of other vehicle components and their feature values. This approach is closest to what has been observed in the data, and therefore will provide individually tailored pricing, and hence more accuracy. No post-hoc RPE or ICM adjustment will be needed.
\end{enumerate}

\subsection{Expert Evaluation}
In addition to the traditional data-driven model validation, the authors attempted to compare the resulting Shapley attribution component price values with existing literature. While the comparison and validation exercise was fruitful and encouraging, literature-available component \textit{cost} (not price) data is at the manufacturing cost level, so the difficulty of a fair comparison was threefold:
\begin{enumerate}
  \item  Component cost is mapped to component pricing through often unreliable RPM and ICM adjustments.
  \item Component cost at the manufacturing level fails to account for interactions and component packaging. This can dramatically affect final pricing.
  \item Component cost values usually neglect to differentiate costing of components by vehicle size, class or powertrain.
\end{enumerate}
The authors performed component price \emph{pseudo} validation in an honest attempt, but were limited as to data availability, data comparability, and knowledge of the field of vehicle/component pricing and the marketing strategies involved. The results of the validation process will be published in a separate article in the future. Meanwhile, we plan to engage in further literature investigation, complete additional analysis and comparisons, attempt to gather more market level component data, reach out to marketing and financial experts, present current method and results outcomes to stakeholders, and, as a result, produce a more comprehensive, engineer-based, validation. We also encourage interested parties and experts to reach out, adopt the methodology and the resulting outcomes and provide suggestions or directions.

\subsection{Model Improvement}
While efforts have been made in the vehicle price modeling to reach $\sim~\$1000$ of average prediction error (equivalent to an average error of 2.2\% of predicted vehicle price), more work can be done towards model error improvement. From the current top-down approach suggested, total vehicle price estimation is used as a basis from which to derive component level prices, i.e., component prices may be affected by vehicles with low prediction accuracy, especially for low-price components. In this spirit, we are encouraged to maintain and continue the modeling exercise.

\subsection{Uncertainty Estimation}
The current suggested approach relies on the fair decomposition of a total vehicle price onto the different component parts using additive feature attribution methods. While the method has certain theoretical guarantees for fairness and optimality, it does not address the uncertainty implicit in the method's outputs. We suggest further investigations of attributional outcome uncertainties, i.e., introducing confidence intervals to quantify the uncertainty in estimated attributions. In return, on the one hand, this will allow us to better quantify how confident one should be about a certain attributed component price for a particular vehicle, and, on the other hand, this will also allow us to exclude or identify uncertain decompositions when deriving global component trends (overall \$/technology feature value).

\section{Data and Code Availability}
The data used in this study was retrieved from several sources of publicly available data. After a substantial effort of data collection process development, data cleaning, data integration and data analysis, the resulting processed \emph{aggregated} data are Argonne property. The database is managed by a MongoDB database management system. The code implementation of the web-scraping process, the clustering, the vehicle price modeling, the Shapley attribution and the data analysis are done in a combination of R, Python, Tableau and Gephi software. For questions and inquiries please contact Ayman Moawad amoawad@anl.gov.

\section*{Acknowledgment}
The authors would like to acknowledge the financial support of Jacob Ward and Heather Croteau (Office of Vehicle Technologies, U.S. Department of Energy) for this work. The submitted manuscript has been created by UChicago Argonne, LLC, Operator of Argonne National Laboratory (Argonne). Argonne, a U.S. Department of Energy Office of Science laboratory, is operated under Contract No. DE-AC02-06CH11357. The U.S. Government retains for itself, and others acting on its behalf, a paid-up nonexclusive, irrevocable worldwide license in said article to reproduce, prepare derivative works, distribute copies to the public, and perform publicly and display publicly, by or on behalf of the Government.\par
The views and opinions of the authors expressed herein do not necessarily state or reflect those of the United States Government or any agency thereof. Neither the United States Government nor any agency thereof, nor any of their employees, makes any warranty, expressed or implied, or assumes any legal liability or responsibility for the accuracy, completeness, or usefulness of any information, apparatus, product, or process disclosed, or represents that its use would not infringe privately owned rights.

\bibliographystyle{IEEEtran}
\bibliography{reference}

\newpage
\begin{IEEEbiographynophoto}{Ayman Moawad}
is a research engineer in the Vehicle and Mobility Simulation group at Argonne National Laboratory. He received a Master's degree in Mechatronics, Robotics, and Computer Science from the Ecole des Mines, France and a Master's degree in Statistics from the University of Chicago, USA. His research interests include engineering applications of artificial intelligence for energy consumption and cost prediction of advanced vehicles, machine learning, large scale data analysis, and high performance computing.
\end{IEEEbiographynophoto}

\begin{IEEEbiographynophoto}{Ehsan Islam}
completed his M. Sc. in Interdisciplinary Engineering from Purdue University, USA in 2019 and B.A.Sc in Mechatronics Engineering from University of Waterloo, Canada in 2016. His skills set and interests focus on applying Mechatronics principles to innovate systems and processes in advanced vehicle technologies and controls systems. At Argonne, he focuses his research on vehicle energy consumption analyses and inputs for U.S. DOE-VTO and NHTSA/EPA/U.S. DOT CAFE and CO$_{2}$ standards using innovative large scale simulation processes and applications of AI.
\end{IEEEbiographynophoto}

\begin{IEEEbiographynophoto}{Namdoo Kim}
graduated in 2007 from the University of Sungkyunkwan, Korea, with a Master’s Degree in Mechanical Engineering. He is currently working in Argonne National Laboratory’s Vehicle Modeling and Simulation Group.
\end{IEEEbiographynophoto}

\begin{IEEEbiographynophoto}{Ram Vijayagopal}
is the technical manager for Vehicle Technology Assessment at Argonne National Laboratory. He is responsible for quantifying the energy saving potential of technologies using modelling and simulation. After working at Mahindra \& Mahindra and Hitachi Automotive Systems, he joined Argonne in 2008. He received his bachelor’s degree in engineering from University of Kerala  and a master’s degree in engineering from University of Michigan. He has authored over 20 papers in the area of advanced vehicle technologies.
\end{IEEEbiographynophoto}

\begin{IEEEbiographynophoto}{Aymeric Rousseau}
is the Manager of the Vehicle and Mobility Simulation group at Argonne National Laboratory. He received his engineering diploma at the Industrial System Engineering School in La Rochelle, France in 1997 and an Executive MBA from Chicago Booth in 2019. For the past 20 years, he has been evaluating the impact of advanced vehicle and transportation technologies from a mobility and energy point of view including the development of Autonomie (vehicle system simulation) and POLARIS (large-scale transportation system simulation). In addition to leading numerous projects for OEMs and US Department of Energy, Aymeric Rousseau has been providing the vehicle energy consumption analysis for the US CAFE regulations and has authored or co-authored more than 80 papers related to vehicle electrification.
\end{IEEEbiographynophoto}

\begin{IEEEbiographynophoto}{Wei Biao Wu}
received his Ph.D. degree in statistics in 2001 from the University of Michigan, Ann Arbor, MI. He is currently a Professor in the Department of Statistics at the University of Chicago, USA.
His research interests include probability theory, statistics, financial time series and econometrics. He is currently interested in high dimensional time series analysis, model selection, and multiple testing problems for dependent data.
\end{IEEEbiographynophoto}

\end{document}